\documentclass[lettersize,journal]{IEEEtran}
\usepackage{array}
\usepackage[caption=false,font=footnotesize,labelfont=rm,textfont=rm,subrefformat=parens,labelformat=parens]{subfig}
\usepackage{textcomp}
\usepackage{stfloats}
\usepackage{url}
\usepackage{verbatim}
\usepackage{graphicx}
\usepackage{cite}
\usepackage{amsmath,amssymb,amsfonts}
\usepackage{xcolor}
\usepackage{CJK}
\usepackage{booktabs}
\usepackage{steinmetz}
\usepackage[linesnumbered,ruled]{algorithm2e}
\usepackage{bm}
\usepackage{makecell}
\usepackage{multirow}
\usepackage{colortbl}
\usepackage[colorlinks=true, linkcolor=blue, citecolor=blue,urlcolor=black]{hyperref}

\begin{document}

\title{Joint Visible Light and Backscatter Communications\\ for Proximity-Based Indoor Asset Tracking \\ Enabled by Energy-Neutral Devices}

\author{Boxuan Xie,~\IEEEmembership{Graduate Student Member,~IEEE},
Lauri Mela,
Alexis A. Dowhuszko,~\IEEEmembership{Senior Member,~IEEE},\\
Yu Bai,~\IEEEmembership{Member,~IEEE},
Zehui Xiong,~\IEEEmembership{Senior Member,~IEEE},
Zhu Han,~\IEEEmembership{Fellow,~IEEE},\\
and Riku Jäntti,~\IEEEmembership{Senior Member,~IEEE}

\thanks{B. Xie, L. Mela, A. Dowhuszko, Y. Bai, and R. Jäntti are with the Department of Information and Communications Engineering, Aalto University, 02150 Espoo, Finland (e-mail: boxuan.xie@aalto.fi, lauri.mela@aalto.fi, alexis.dowhuszko@aalto.fi, yu.bai@aalto.fi, riku.jantti@aalto.fi)}

\thanks{Z. Xiong is with the School of Electronics, Electrical Engineering and Computer Science (EEECS), Queen's University Belfast, Belfast, BT7 1NN, U.K. (z.xiong@qub.ac.uk).}

\thanks{Z. Han is with the Department of Electrical and Computer Engineering, University of Houston, Houston, TX 77004 USA (email: hanzhu22@gmail.com).}

\vspace{-20pt}
}

\markboth{Journal of \LaTeX\ Class Files,~Vol.~14, No.~8, August~2021}%
{Xie \MakeLowercase{\textit{et al.}}: Joint Visible Light and Backscatter Communications for Indoor Asset Tracking}

\maketitle
\begin{abstract}
In next-generation wireless systems, providing location-based mobile computing services for energy-neutral devices has become a crucial objective for a sustainable Internet of Things~(IoT). 
Visible light positioning~(VLP) has attracted substantial research attention as a complementary method to radio frequency (RF) solutions since it can leverage ubiquitous lighting infrastructure. However, conventional VLP receivers often rely on photodetectors or cameras that are power-hungry, complex, and expensive.
To address this challenge, we propose a hybrid indoor asset tracking system that integrates visible light communication~(VLC) and backscatter communication~(BC) within a simultaneous lightwave information and power transfer~(SLIPT) framework.
We design a low-complexity and energy-neutral IoT node, namely a backscatter device~(BD), which harvests energy from light-emitting diode (LED) access points, and then modulates and reflects ambient RF carriers to indicate its location within particular VLC cells. 
We present a multi-cell VLC deployment using frequency-division multiplexing (FDM) that mitigates interference among LED access points by assigning them distinct frequency pairs based on a four-color map scheduling principle. 
We develop a lightweight particle filter~(PF) tracking algorithm at an edge RF reader. The algorithm fuses proximity reports with the received backscatter signal strength to track the BD. 
Experimental results show that this approach achieves positioning errors of 0.318~m at the 50th percentile and 0.634~m at the 90th percentile,
while avoiding the use of complex photodetectors and an active RF synthesizer at the energy-neutral IoT node. 
These results demonstrate the feasibility of energy-neutral, edge-assisted indoor tracking with a low-complexity device across the tested trajectories.
\end{abstract}

\vspace{-10pt}
\section{Introduction}\label{sec:intro}
\IEEEPARstart{I}{n} the emerging sixth-generation~(6G) era, growing demand for energy-efficient and cost-effective Internet of Things~(IoT) devices aligns with future network key impact factors~(KPIs) of reduced power consumption and complexity~\cite{lopez2025zed,saad20206g_survey}.
Meanwhile, many IoT devices require accurate indoor positioning services, especially for application scenarios such as logistics, healthcare, and industrial automation. 
These services rely on seamless localization and tracking of IoT devices, which are challenging in multipath-rich indoor environments~\cite{farahsari2022pos_survey}. 
A recent proposal within the Third Generation Partnership Project~(3GPP), referred to as Ambient IoT~\cite{3gpp38848}, highlights energy-neutral IoT nodes with built-in positioning functionalities, drawing further attention to low-power, low-cost, and high-accuracy indoor positioning methods~\cite{kimionis2024aiot}. This increasing interest underlines the significance of exploring new solutions that integrate efficient energy harvesting and localization capabilities for IoT devices.

Global Navigation Satellite Systems (GNSSs) offer reliable positioning in outdoor areas~\cite{gnss2023survey}, yet the positioning accuracy degrades inside buildings due to severe signal attenuation and multipath fading.
As a result, a variety of radio frequency (RF) technologies, including Wi-Fi~\cite{wu2020wifi}, Bluetooth~\cite{faragher2015bt_pos}, LoRa~\cite{hu2022tracking_lora}, cellular networks~\cite{zhou2024pos_5g}, radio frequency identification (RFID)~\cite{xu2023rfid_survey}, and ultra-wideband (UWB)~\cite{guo2024tracking_uwb}, have been investigated for indoor positioning. 
These systems exploit measurements of received signal strength (RSS), time of arrival (ToA), time difference of arrival (TDoA), or angle of arrival (AoA), which can be augmented by multi-antenna techniques to improve the positioning accuracy. 
However, the inherent trade-offs between positioning accuracy, hardware complexity, and deployment cost remain practical constraints for low-complexity IoT implementations.

The broad adoption of light-emitting diodes (LEDs) has increased interest in visible light communication~(VLC), enabling luminaires to transmit information over optical wireless links~\cite{haas2019survey}. 
VLC offers a promising pathway for indoor localization, typically referred to as visible light positioning~(VLP)~\cite{haas2018vlp_survey,zhu2024vlp_survey,bastiaens2024vlp_survey}. 
Previous works by Zhu~\textit{et al.}~\cite{zhu2024vlp_survey} and Bastiaens~\textit{et al.}~\cite{bastiaens2024vlp_survey} have presented comprehensive overviews of VLP fundamentals and system designs. 
Studies by Yang~\textit{et al.}~\cite{yang2020vlp_survey,yang2023vlp_advance,yang2020vlp_resource} investigated VLP using RSS-, phase-, and time-based measurements to accommodate various positioning accuracy requirements. 
They have also investigated resource allocation to provide different quality-of-service (QoS) levels in VLP systems~\cite{yang2020vlp_qos,yang2020vlp_resource}, aiming to ensure stable indoor communication and positioning for IoT nodes. 
Despite these advances, VLC receivers in such systems often rely on specialized photodetectors or camera sensors and incorporate energy-intensive components, such as transimpedance amplifiers~(TIAs), resulting in increased power consumption and hardware complexity, which in turn adds to the overall cost on the IoT device side.
Therefore, designing low-power and low-complexity VLP solutions remains a key challenge for cost-effective IoT deployment.

Among VLP technologies, proximity-based positioning is one simple method to localize IoT devices under indoor illumination, as the location of the IoT device is approximated by the known position of the nearest VLC access points~(APs)~\cite{guillermo2013prox,cherntanomwong2015prox,xie2018vlp_proximity}. 
Although these proximity-based approaches suit many low-data-rate scenarios, they assume that the IoT device performs both VLC reception and location estimation. This dual role is less practical for asset tracking, where position reports are usually collected and monitored by a distant edge reader.
Moreover, conventional VLC receivers rely on specialized transmission and reception hardware, which limits their applicability in resource-limited and energy-constrained IoT devices. Therefore, there is a need for an energy-efficient and low-complexity VLC-based proximity-reporting method for IoT devices to enable scalable indoor IoT asset tracking.

Besides conventional camera-, photodiode-, and APD-based VLP systems, recent literature has also started to explore self-powered or energy-harvesting-assisted light-based positioning, where photovoltaic modules can be used not only as optical receivers but also as energy sources for low-power positioning nodes~\cite{cappelli2022selfsufficient,perera2025mlaided,zhou2025resource}. 
Retroreflector-based passive VLP has also emerged as a related direction, where both the downlink and uplink directions of communication operate in the optical domain~\cite{li2015retro,shao2018retro,shao2019passiveretro, nazzal2022retrovlp}.
However, existing self-powered or retroreflector-based VLP studies typically perform on-node localization computations in the optical domain, focus on static positioning or orientation estimation, and rely on active-radio or optical mechanisms to report localization information.

Recently, hybrid VLC/RF networks have emerged as a promising strategy for combining the high data rates of VLC with the ubiquitous coverage of RF systems~\cite{han2022survey,abuella2021hybrid_survey}. 
Early studies focused on IoT devices equipped with both VLC receivers and active RF transmitters to relay VLC data to conventional RF receivers~\cite{letaief2021hybrid, peng2021end}.
Alongside these relay-based approaches, simultaneous lightwave information and power transfer~(SLIPT) has gained momentum as a technique to energize IoT devices via LED illumination while delivering data~\cite{ding2018slipt,xiao2021slipt}.
For instance, Peng~\textit{et al.} considered optimizing the end-to-end performance of an IoT relay-enabled hybrid VLC/RF system under transmission-power constraints~\cite{peng2021end}, and Tang~\textit{et al.} studied sustainability-driven resource-allocation strategies for hybrid VLC/RF systems~\cite{xiong2024hybrid2}. 
Although these studies have shown the potential to achieve lower operating costs and improved energy efficiency, they generally assume active RF transmitters on the IoT nodes, which increases device complexity.

To address the challenge of reporting proximity information through RF channels without active RF components, recent advances in ambient backscatter communication (BC) provide an alternative that allows ultra-low-power RF transmissions by modulating existing RF signals without requiring energy-intensive local oscillators and power amplifiers on IoT devices~\cite{liu2013ambient,van2018survey, jiang2023survey}.
In this regard, communications schemes integrating VLC with BC in hybrid VLC/RF networks have been explored in~\cite{varshney2018vlbc,varshney2023tunnellifi,varshney2024lifibc,xie2024vlc,xie2024light}.
Among these works, Giustiniano~\textit{et al.} introduced a system named BackVLC that employs LED bulbs for both illumination and data transmission~\cite{varshney2018vlbc}, enabling battery-free IoT tags to communicate using visible light and RF backscatter signals.
Similarly, Mir~\textit{et al.} proposed TunnelLiFi~\cite{varshney2023tunnellifi}, an IoT node that bridges visible light and radio spectrum using a tunnel diode oscillator for low-power conversion, enabling the relaying of LiFi signals to other IoT devices. 
They also developed PassiveLiFi~\cite{varshney2024lifibc}, a system integrating VLC with RF backscatter communications, which uses a chirp spread spectrum VLC transmitter to improve the communication range and efficiency.
Xie~\textit{et al.} explored LiBD~\cite{xie2024vlc}, a backscatter device (BD) that uses visible light for both control and power purposes in BC over sub-6~GHz bands. They also designed a thin-film BD fabricated through inkjet printing and additive manufacturing on flexible substrates~\cite{xie2024light}, which converts light-carrying data into RF signals and operates under different lighting conditions.
Koskinen~\textit{et al.} developed Li2BC~\cite{koskinen2025li2bc}, a BD that integrates VLC reception and processing with backscatter modulation using a low-power microcontroller. 

Despite these developments, one specific problem remains insufficiently explored: how to realize indoor asset tracking in an Ambient IoT scenario with an energy-neutral and low-complexity device that can receive location information from LED luminaires and report it to a remote reader without relying on complicated VLC receivers or active RF transceivers. 
This challenge is very relevant for asset tracking applications, where the tracked device should be inexpensive, maintenance-free, and lightweight, while still providing practically useful positioning accuracy.
The key design challenge is to transfer proximity information generated in the optical domain to an RF-domain edge reader, where the information processing to perform the target asset tracking application can be performed. 
The existing proximity-based VLP systems assume that VLC reception and position estimation are both performed on the device side, while existing hybrid VLC/RF solutions often rely on active RF transmission at the IoT node.

To address this problem, we propose a joint VLC-BC-enabled indoor asset tracking system. The design features a multi-cell, frequency-division multiplexing~(FDM)-based VLC transmission system in which multiple LED luminaires illuminate the service area and broadcast location-specific identifiers to a BD as proximity information. We further design a low-complexity, battery-free BD that harvests energy from the VLC signals and modulates ambient RF carrier waves to report the received proximity information to an edge reader, where a tracking algorithm estimates the position of the BD. 
The key contributions of this paper are listed as follows:
\begin{itemize}
    \item
    We propose an end-to-end indoor asset-tracking architecture that integrates VLC and BC within a SLIPT framework. In this architecture, a battery-free BD acts as a light-to-RF relay; that is, it harvests optical energy from LED illumination, receives LED-cell proximity information through a VLC channel, and reports this information to an edge reader by passive RF backscatter communication channel. This design shifts communication and localization complexity away from the IoT node, providing a device-side energy-neutral and low-complexity operating point for indoor tracking.

    \item
    We design a multi-cell VLC proximity-reporting scheme based on FDM to support concurrent LED-ID transmission from multiple luminaires. Each LED AP embeds its ID using binary frequency-shift keying (BFSK) modulation, while adjacent VLC cells are assigned distinct frequency pairs following a four-color-inspired frequency-reuse principle. This scheme mitigates strong interference among neighboring cells and enables the BD to distinguish single-cell and overlapping-cell illumination regions without requiring tight time synchronization.

    \item
    We develop a particle filter (PF)-based tracking algorithm at the edge RF reader to estimate the BD position from different types of  measurements. The algorithm combines discrete proximity reports, represented by the decoded LED-ID set, with continuous RSS measurements from the BC link. This fusion allows VLC proximity information to constrain the feasible tracking region, while the BC RSS observation refines the position estimate under measurement noise and indoor multipath effects.

    \item
    We implement a proof-of-concept system that demonstrates the complete VLC-to-BC tracking chain among LED luminaires, a battery-free BD, an RFS, and an RF reader. Under the tested conditions, the prototype shows that the BD can harvest light energy, receive LED IDs, and backscatter proximity information without active RF syntisizers. Simulations and experiments over multiple trajectories yield submeter-level positioning accuracy, indicating the potential of the proposed system to implement low-complexity indoor asset-tracking applications.
\end{itemize}

The rest of the paper is organized as follows. Section~\ref{sec:system_model} introduces the joint VLC-BC system model. 
Section~\ref{sec:problem_tracking} describes the tracking task, presents the details of the multi-cell VLC deployment, and describes the PF-based tracking algorithm design.
Section~\ref{sec:implementations} illustrates
the implementation of the proposed system with end-to-end communication modules.
Then, Section~\ref{sec:sim_exp_results} presents and evaluates the simulation and experimental results.
Finally, Section~\ref{sec:conclusion} draws conclusions.
\section{System Model}\label{sec:system_model}
\begin{figure}
\centering
\includegraphics[width=0.98\columnwidth]{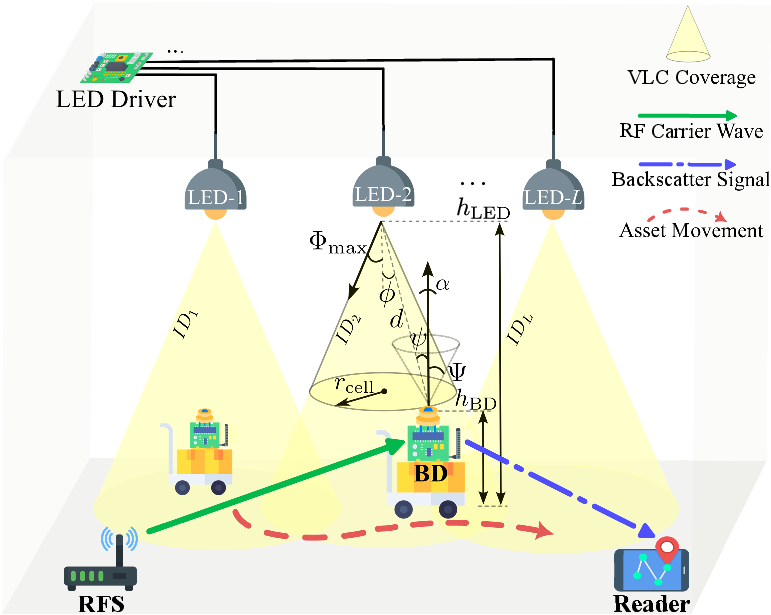}
\caption{Overview of the joint VLC-BC system proposed for indoor asset tracking. The BD attached to the asset receives LED IDs over light (yellow cones) and converts them into an electrical signal to modulate the ambient RF carrier waves (green arrow) emitted by the RFS. The modulated signal is backscattered (blue arrow) and detected by an edge reader for location inference.}
\label{fig:system_model_integrate}
\end{figure}
The proposed joint VLC-BC system for indoor asset tracking is depicted in Fig.~\ref{fig:system_model_integrate}.
In this paper, we consider a single-target scenario, where the position of one asset moving across the floor is tracked with the aid of an edge reader. One IoT node, namely a BD, is attached directly to that asset.
Multiple LED APs illuminate the service area and continuously broadcast their unique IDs through VLC. Each of these IDs can be mapped to the known location of an LED AP in a database.
When the BD moves into the coverage area of an LED AP, it receives the ID of the LED AP from the VLC link, and the BD forwards that ID to the distant reader by modulating and backscattering ambient RF carriers emitted by an RFS over the BC link.
Finally, the reader captures the backscattered signal and decodes the LED IDs, which enables it to estimate the position of the BD.
Successful operation of this system requires the simultaneous availability of two resources at the BD, namely sufficient optical illumination for energy harvesting and VLC reception, and sufficient RF carrier illumination for backscatter modulation. If either the optical input or the RF carrier becomes too weak, the quality of the end-to-end report degrades and may eventually lead to a temporary outage of the tracking link.

\begin{figure*}
\centering
\includegraphics[width=1.98\columnwidth]{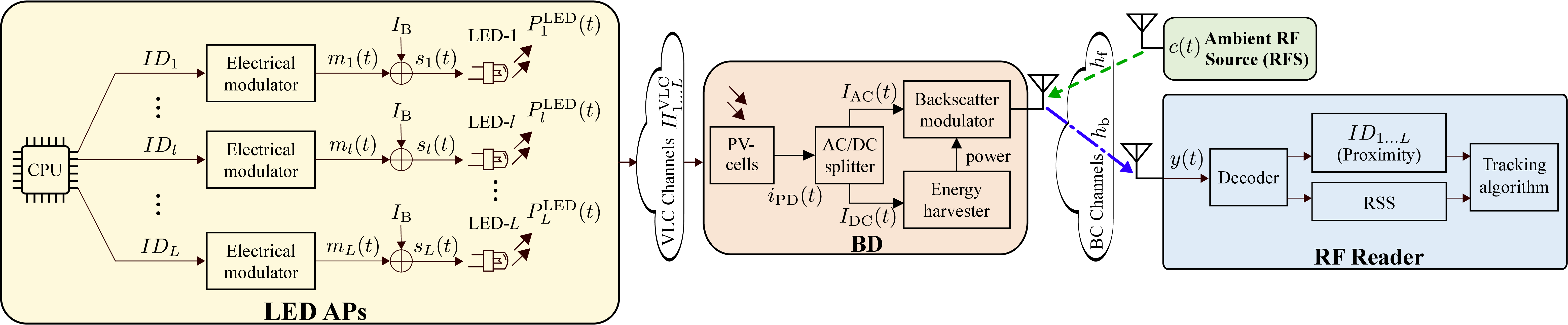}
\caption{A schematic diagram of the joint VLC-BC system. LED APs illuminate the indoor space and transmit unique IDs through VLC links. The BD receives VLC signals and converts them into electrical signals, whose DC components are used to power the BD while AC components are used to modulate the captured RF carriers emitted by the RFS. The modulated signal is backscattered toward the RF reader, where it is decoded and used for positioning purposes.}
\label{fig:system_schematic}
\end{figure*}
The detailed schematic diagram of the system is presented in Fig.~\ref{fig:system_schematic}. 
LED APs transmit VLC signals to the BD using an intensity modulation/direct detection~(IM/DD) scheme, and the BD receives their messages when it is within the coverage range of the LED APs.
The BD in this setup consists of a photodetector, an energy harvester, a backscatter modulator, and an RF antenna~\cite{xie2024vlc, xie2024light}.
Since the LED APs are designed to transmit their IDs at low data rates on the order of a few kilobits per second, this paper considers the use of photovoltaic~(PV) cells as photodetectors. 
The alternating current/direct current~(AC/DC) splitter separates the photovoltaic-converted signals into electrical AC and DC components. The AC component, which carries the VLC message, is used to control the backscatter modulation. 
The DC component supplies energy to the BD.
The BD then modulates the captured ambient RF carrier waves coming from the RFS, which can be, for example, a WLAN AP. 
The AC component controls the RF carrier amplitude by generating the required alternation between the reflection coefficients of the BD.
The modulated signal is then backscattered and captured by the reader equipped with conventional RF front ends. The reader decodes the signal containing the proximity information of the BD and measures the RSS of the received signal. These two measurements are combined by a tracking algorithm to localize and track the position of the BD.
In this configuration, the BD itself does not perform LED-ID decoding, packet synchronization, or localization. Instead, the BD only converts the incident light waveform into a switching signal for RF backscatter modulation, whereas the reader performs synchronization, decodes the LED IDs contained in the backscattered signal, measures the RSS, and combines these observations for localization and tracking.
The following subsections briefly describe the VLC and BC links used in this paper.
\subsection{VLC Channel Model}\label{subsec:vlc_channel}
In the downlink direction of the VLC transmission, each LED AP is assigned a unique ID and transmits its individual VLC signal using FDM. 
BFSK baseband modulation is used to generate the square-wave AC current that modulates the intensity of the light emitted by the LED, while a bias-tee is employed to add a DC bias to the modulating signal.
Let $m(t)$ denote the LED-modulating AC signal at time~$t$, which can be expressed by
\begin{equation}
m(t) = 
\begin{cases} 
 \text{sgn}\left(\sin(2 \pi f_1 t)\right), & \text{bit} = 0, \\
 \text{sgn}\left(\sin(2 \pi f_2 t)\right), & \text{bit} = 1,
\end{cases}
\end{equation}
where $\text{sgn}(\cdot)$ denotes the sign function and $f_{1}$ and $f_{2}$ represent the modulating frequencies of the BFSK scheme. 
Let $l=1, \dots, L$ represent the indices of LED APs.
Then, the electrical signal that drives the $l$-th LED AP can be written as $s_{l}(t) = m_{l}(t) + I_{\textrm{B}}$, where $I_{\textrm{B}}$ represents the DC biasing current driving the LED.
Therefore, the transmitted optical signal can be expressed by~\cite{ding2018slipt}
\begin{equation}
    P_{l}^{\textrm{LED}}(t) =  \eta_{\textrm{E-O}} \left( m_{l}(t) + I_{\textrm{B}} \right),
\end{equation} 
where $\eta_{\textrm{E-O}}$ is the LED electric-to-optical power conversion factor.
Since a practical LED transmitter operates within a limited linear region, the peak amplitude $s_{\textrm{max}}$ of $s_{l}(t)$ should be bounded to avoid clipping. Specifically, 
\mbox{$s_{\textrm{max}} \leq \textrm{min} \left( I_{\textrm{B}} - I_{\textrm{min}}, I_{\textrm{max}} - I_{\textrm{B}} \right),$}
where $I_{\textrm{max}}$ and $I_{\textrm{min}}$ denote the maximum and minimum driving current, respectively. 

To characterize the VLC channel, we adopt the direct illumination model shown in Fig.~\ref{fig:system_model_integrate}, assuming that the direct optical signal is significantly stronger than any reflected signals~\cite{alexis2024planning}. 
Phosphor-coated LEDs are considered and assumed to emit light following a Lambertian radiation pattern. Within its coverage area, each LED AP maintains a line-of-sight~(LoS) link to the BD.
This modeling implies an inherent sensitivity to optical blockage. In particular, if the direct optical paths from all illuminating LED APs to the BD are blocked by an obstacle for a sufficiently long duration, the BD may temporarily lose the harvested optical energy and/or the LED-cell information.
For simplicity, we assume that the BD and its photodetector share the same position. A prototype is shown in Fig.~\ref{fig:bd}.
Then, the DC gain of the optical channel between $\textrm{LED-}l$ and the BD-equipped photodetector can be written as~\cite{alexis2024planning}
\begin{equation}
\label{eqn:vlc_gain}
H_{l}^{\textrm{VLC}} =
\begin{cases}
\frac{(\nu+1)A_{\textrm{PD}}}{2\pi d_{l}^2} \cos^\nu(\phi_{l}) \cos(\psi_{l}), & |\psi_{l}| \leq \Psi, \\
0, & |\psi_{l}| > \Psi,
\end{cases}
\end{equation}
where $d_{l}$ denotes the Euclidean distance between $\textrm{LED-}l$ and the BD, and $\nu = -1 / \log_2 [\cos(\Phi_{\textrm{max}})]$ denotes the Lambertian index of the LEDs. 
The parameter $\Phi_{\textrm{max}}$ indicates the LED half-power semi-angle, whereas $\Psi$ defines the field-of-view~(FoV) semi-angle of the photodetector with an active area of $A_{\textrm{PD}}$. Furthermore, $\phi_{l} \geq 0$ and $\psi_{l}$ are the irradiance and incidence angles of the LoS link between $\textrm{LED-}l$ and the BD, respectively.
The BD-orientation angle $\alpha_{l} = \phi_{l}-\psi_{l}$ represents the angular deviation of the photodetector's acceptance-cone axis from the upward direction.
This dependence on $\psi_l$ also implies sensitivity to the physical orientation of the BD. If the PV-cell receiver is tilted away from the ceiling, mounted sideways, or turned upside down, the effective optical channel gain would likely decrease since the incidence angle moves away from the direction in which the axis of the FoV acceptance cone of the solar panel is pointing. In the extreme case, the received optical signal may fall below the level required for reliable ID recovery and/or sufficient energy harvesting for operation.

The aggregated VLC signal power that arrives at the BD can be expressed by
$\Sigma_{l=1}^{L} P_{l}^{\textrm{LED}} H_{l}^{\textrm{VLC}}$, where $L$ is the total number of LED APs.
Therefore, the BD-equipped photodetector produces an electrical current given by
\begin{equation}
\begin{split}
    i_{\textrm{PD}}(t) &= \eta_{\textrm{O-E}} \left( 
    \Sigma_{l=1}^{L} P_{l}^{\textrm{LED}} H_{l}^{\textrm{VLC}} \right) + n(t)\\
    &= I_{\textrm{AC}}(t) + I_{\textrm{DC}}(t) + n(t),
\end{split}
\end{equation}
where $\eta_{\textrm{O-E}}$ is the optical-to-electrical responsivity of the photodetector, $I_{\textrm{AC}}(t)$ and $I_{\textrm{DC}}(t)$ are the AC and DC components of the received optical signal, respectively, and $n(t)$ is additive white Gaussian noise with variance $N_{0}B$, with $N_{0}$ being the noise power spectral density and $B$ the system operational bandwidth. 
Furthermore, the output AC component carrying the VLC message is given by
\begin{equation}
\label{eqn:i_ac}
    I_{\textrm{AC}}(t) = \eta_{\textrm{O-E}} \Sigma_{l=1}^{L} \eta_{\textrm{E-O}}  H_{l}^{\textrm{VLC}} m_{l}(t).
\end{equation}
The BD harvests energy for its circuitry, including the backscatter modulator, from the DC component $I_{\textrm{DC}}(t)$ output by the BD-equipped photodetector. The amount of energy that can be harvested from the received VLC signal can be written as~\cite{letaief2021hybrid}
\begin{equation}
\label{eqn:eh}
    E_{\textrm{h}} = \varepsilon I_{\textrm{DC}} V_{\textrm{OC}} = \varepsilon \eta_{\textrm{O-E}} \left(\Sigma_{l=1}^{L} \eta_{\textrm{E-O}}  
    H_{l}^{\textrm{VLC}} I_{\textrm{B}} \right) V_{\textrm{OC}},
\end{equation}
where $\varepsilon$ is the fill factor of the photodetector, $V_{\textrm{OC}}$ is the open-circuit voltage given by
\begin{equation}
    V_{\textrm{OC}} = V_{\textrm{t}} \ln \left( 1 + \frac{\eta_{\textrm{O-E}} \left( \Sigma_{l=1}^{L} \eta_{\textrm{E-O}} 
    H_{l}^{\textrm{VLC}} I_{\textrm{B}} \right)}{I_0} \right),
\end{equation}
where $V_{\textrm{t}}$ is the thermal voltage and $I_{0}$ is the dark saturation current of the BD-equipped photodetector.
\subsection{Backscatter Channel Model}\label{subsec:bc_channel}
The BC link comprises an RFS, a BD, and an RF reader. The BD modulates the carrier signals emitted by the RFS using photovoltaic-converted VLC signals and subsequently backscatters the modulated signals to the reader.
In this way, the system shifts the radio transmission burden from the IoT node to a shared carrier-emitting infrastructure.
The received backscatter signal at the reader is then expressed by~\cite{griffin2009linkbudget} 
\begin{equation}
    y(t) = \sqrt{\xi G_{\textrm{T}} G_{\textrm{R}}} G_{\textrm{BD}} h_{\textrm{f}}(t)  h_{\textrm{b}}(t) c(t) I_{\textrm{AC}}(t) + n(t),
\end{equation}
\noindent where $\xi$ denotes the backscatter efficiency of the BD and $G_{\textrm{T}}$, $G_{\textrm{R}}$, and $G_{\textrm{BD}}$ are the antenna gains of the RFS, reader, and BD, respectively.
The symbol $h_{\textrm{f}}$ is the forward channel gain (RFS$\rightarrow$BD) between the RFS and the BD, and $h_{\textrm{b}}$ is the backscatter channel gain (BD$\rightarrow$reader) between the BD and the reader.
The ambient RF carrier signal $c(t)$ coming from the RFS has a mean power of $ P_{\textrm{c}} = \mathbb{E} \{ \left| c(t) \right|^2 \}$. 
In a practical deployment based on ambient RF infrastructure, the effective carrier level may fluctuate over time due to traffic dynamics, transmit-power variation, or channel changes. In this case, the absolute RSS of the backscatter signal is no longer influenced only by geometry and tag state, but also by temporal variations of the carrier source.
Furthermore, $I_{\textrm{AC}}(t)$ represents the AC component of the photovoltaic-converted VLC signal presented in~\eqref{eqn:i_ac} and is used to modulate the RF carrier signal.
The term $n(t)$ is additive white Gaussian noise with variance $N_{0} B$.
Moreover, $\xi$ is related to the polarization mismatch $\chi_\textrm{f}$ between the RFS and BD, and $\chi_\textrm{b}$ models the mismatch between the BD and reader. 
Finally, the modulation factor of the BD is given by $M$, and $\Theta$ represents an object penalty; thus, the backscatter efficiency can be expressed by \mbox{$\xi = (\chi_\textrm{f} \chi_\textrm{b}M)/ \Theta^{2}$}~\cite{griffin2009linkbudget}.
For practical reasons, the radio channels $h_{\textrm{f}}$ and $h_{\textrm{b}}$ are modeled using the 3GPP Indoor Hotspot model~\cite{3gpp38901}. 
The model defines path losses for LoS and non-line-of-sight (NLoS) conditions, expressed by
\begin{equation}
\begin{split}
\mathcal{L}_{\textrm{LoS,dB}} = 32.4 + 17.3\log_{10}(d_{\textrm{3D}})  + 20\log_{10}(f_{\textrm{c}}) + \delta_\textrm{LoS,dB},
\end{split}
  \label{eq:3GPPInHPathLoss_LoS}
\end{equation}
\begin{equation}
\begin{split}
\mathcal{L}_{\textrm{NLoS,dB}} = \max(\mathcal{L}_{\textrm{LoS,dB}}, 17.3 + 38.3\log_{10}(d_{\textrm{3D}}) \\ + 24.9\log_{10}(f_{\textrm{c}}) + \delta_\textrm{NLoS,dB}),
\end{split}
  \label{eq:3GPPInHPathLoss_NLoS}
\end{equation}
where $d_{\textrm{3D}}$ represents the 3D Euclidean distance between a transmitter and a receiver, $f_{\textrm{c}}$ is the carrier frequency in gigahertz, and $\delta_\textrm{LoS,dB}$ is the shadowing factor modeled as a lognormal random variable with $\sigma_{\textrm{LoS}} = 3$~dB for \mbox{$1 \leq d_{\textrm{3D}} \leq 150$~m}.
Similarly, $\delta_\textrm{NLoS,dB}$ is the shadowing factor modeled as a lognormal random variable with \mbox{$\sigma_{\textrm{NLoS}} = 8.03$~dB} for \mbox{$1 \leq d_{\textrm{3D}} \leq 150$~m}.
The probability of an LoS link depends on the target environment. In an open indoor environment, the LoS probability is given by~\cite{3gpp38901}
\begin{equation} \label{eq:Indoor3GPP_Pr_LoS_Open}
\textrm{Pr}_{\textrm{LoS}} =
\begin{cases}
1, & d_{\textrm{2D}} \leq 5~\textrm{m}, \\
e^{-\frac{d_{\textrm{2D}} - 5}{70.8}}, & 5 < d_{\textrm{2D}} \leq 49~\textrm{m}, \\
0.54 e^{-\frac{d_{\textrm{2D}} - 49}{211.7}}, & d_{\textrm{2D}} > 49~\textrm{m},
\end{cases}
\end{equation}
where $d_{\textrm{2D}}$ is the horizontal distance between the transmitter and receiver.
Finally, the expected value of the overall path loss at a given distance, which combines the LoS and NLoS conditions weighted by their respective probabilities, can be written as
\begin{equation}
\label{eq:3GPPInHPathLoss_Overall}
\mathcal{L}_{\textrm{dB}} = \textrm{Pr}_{\textrm{LoS}}\, \mathcal{L}_{\textrm{LoS,dB}} + (1-\textrm{Pr}_{\textrm{LoS}})\, \mathcal{L}_{\textrm{NLoS,dB}}.
\end{equation}
Hence, the RSS of the received backscatter signal can be expressed by
\begin{equation}
\label{eqn:bc_rss}
    R^\textrm{BC} = \xi G_{\textrm{T}} G_{\textrm{R}}G_{\textrm{BD}}^2 I^2_{\textrm{AC}}
    P_{\textrm{c}} \mathcal{L}_{\textrm{f}}^{-1} \mathcal{L}_{\textrm{b}}^{-1},
\end{equation}
where $\mathcal{L}_{\textrm{f}} = |h_{\textrm{f}}|^{-2} $ and $ \mathcal{L}_{\textrm{b}} = |h_{\textrm{b}}|^{-2}$ represent the path losses of the forward and backscatter channels, respectively.
\section{Proximity-Based Asset Tracking}
\label{sec:problem_tracking}
This section describes the task of tracking the BD in an indoor environment based on the proximity information coming from LED APs and on the RSS of backscatter signals. 
We first present the underlying challenges and detail the tracking task in a state-space framework in Section~\ref{subsec:problem_formulation}. 
We then describe the VLC cell model in Section~\ref{subsec:vlc_cell_model}, where each LED AP provides coverage in a limited region, and explain how the FDM-based approach enables the reception of multiple overlapping light signals in Section~\ref{subsec:vlc_deployment}. Finally, we develop a lightweight particle-filter-based algorithm to fuse the proximity information of the BD and the RSS of the backscatter signal for tracking in Section~\ref{subsec:pf_tracking}.

\subsection{Task Description}
\label{subsec:problem_formulation}
Consider a set of ceiling-mounted LED APs illuminating an indoor environment. In the present formulation, one BD is tracked in the service area at a time.
As the BD moves across this service area, it detects the unique IDs of any nearby LED APs that are in coverage range.
These IDs are then forwarded to a distant RF reader using the backscatter communication mechanism. Consequently, the proximity of the BD to a particular LED AP, or sets of APs with overlapping coverage regions, is reported whenever the BD is illuminated by one or more APs. If the BD is outside all of the VLC coverage areas, it does not receive and forward any ID.
By decoding and analyzing the backscattered signal, the reader can estimate the position of the BD.
To formulate the above mechanism as a tracking problem, we identify three main issues. 
First, we need an accurate representation of the coverage region of an LED AP, also known as a VLC-cell service area, to determine whether the BD is within the coverage range of the cell. 
Second, when moving into overlapping regions of VLC cells, the BD will likely detect multiple VLC signals and multiple LED IDs simultaneously, thus requiring a suitable multiplexing method to distinguish them. 
Third, we need to develop a tracking algorithm at the reader that incorporates two measurements: (i) the set of LED IDs detected and forwarded by the BD, and (ii) the RSS of the backscatter signal, to localize and track the BD over time.

For a multi-BD scenario, the proposed framework would require a lightweight medium access-control (MAC) layer. One option is TDMA, where the reader or the LED infrastructure provides a common frame clock, for example through a dedicated RF beacon or a global optical synchronization pattern, and each BD is assigned a time slot during which its RF switch is enabled. 
Another option is slotted ALOHA, where the same slot structure is used, but each BD selects a transmission slot following a contention-based random process whenever it needs to report a new proximity observation. 
In both cases, the current VLC-based proximity sensing mechanism remains unchanged, while the access-control logic determines the modulation timing of each BD to keep under control the probability of collisions among multiple BDs.
\subsection{VLC Cell Model}
\label{subsec:vlc_cell_model}
The coverage area of each LED AP can be approximated as a circle on the light-receiving plane parallel to the floor, known as a VLC cell service area~\cite{alexis2024planning}.
This approximation is determined by the LoS geometry between the LED AP and the BD-equipped photodetector. 
As illustrated in Fig.~\ref{fig:system_model_integrate}, let $h_{\text{LED}}$ and $h_{\textrm{BD}}$ denote the vertical positions of a ceiling-mounted LED AP and the BD above the floor, respectively. Suppose the BD is oriented toward the ceiling with an FoV semi-angle~$\Psi$. Then, the nominal radius of the VLC cell is expressed by~\cite{alexis2024planning}
\begin{equation}\label{eq:vlc_cell_radius}
    r_\textrm{cell} = (h_{\text{LED}} - h_{\textrm{BD}}) \,\tan(\Psi).
\end{equation}
When the horizontal distance from the BD to an LED AP does not exceed $r_\textrm{cell}$, the BD can receive the VLC signal if the link budget is sufficient. If the BD moves outside the VLC cell radius, the VLC link is disrupted under the adopted geometric model shown in~\eqref{eqn:vlc_gain}. This geometric relation also depends on rotation or tilting of the BD~\cite{alexis2024planning,guvenc2024orientation}; for simplicity, however, this paper considers an upward orientation of the BD.
This nominal circular service area is therefore valid for the simplified study case in which the BD is approximately upward-facing. In practical warehouse or office deployments, however, a tag attached to a box or asset may become tilted, sideways, or inverted. 
Such orientation changes reduce the effective optical gain and distort the nominal service area, which may in turn create additional local outages even if the geometric floor projection of the BD remains inside the ideal VLC cell.
This geometric model also reveals a practical deployment trade-off. If the spacing among LED APs is too large relative to $r_\textrm{cell}$, or if one luminaire becomes unavailable without sufficient overlap from neighboring VLC cells, uncovered regions may eventually appear where the BD cannot reliably receive LED-cell information. In the current design, such regions may also lead to reduced energy harvesting ability in the BD, thereby degrading the continuity of the backscatter communication link. Therefore, a reliable deployment of VLC AP infrastructure requires not only coverage of individual cells, but also adequate overlap and redundancy among neighboring luminaires.
\subsection{Deployment of the Multi-Cell VLC System}
\label{subsec:vlc_deployment}
\begin{figure}
\centering
\includegraphics[width=0.98\columnwidth]{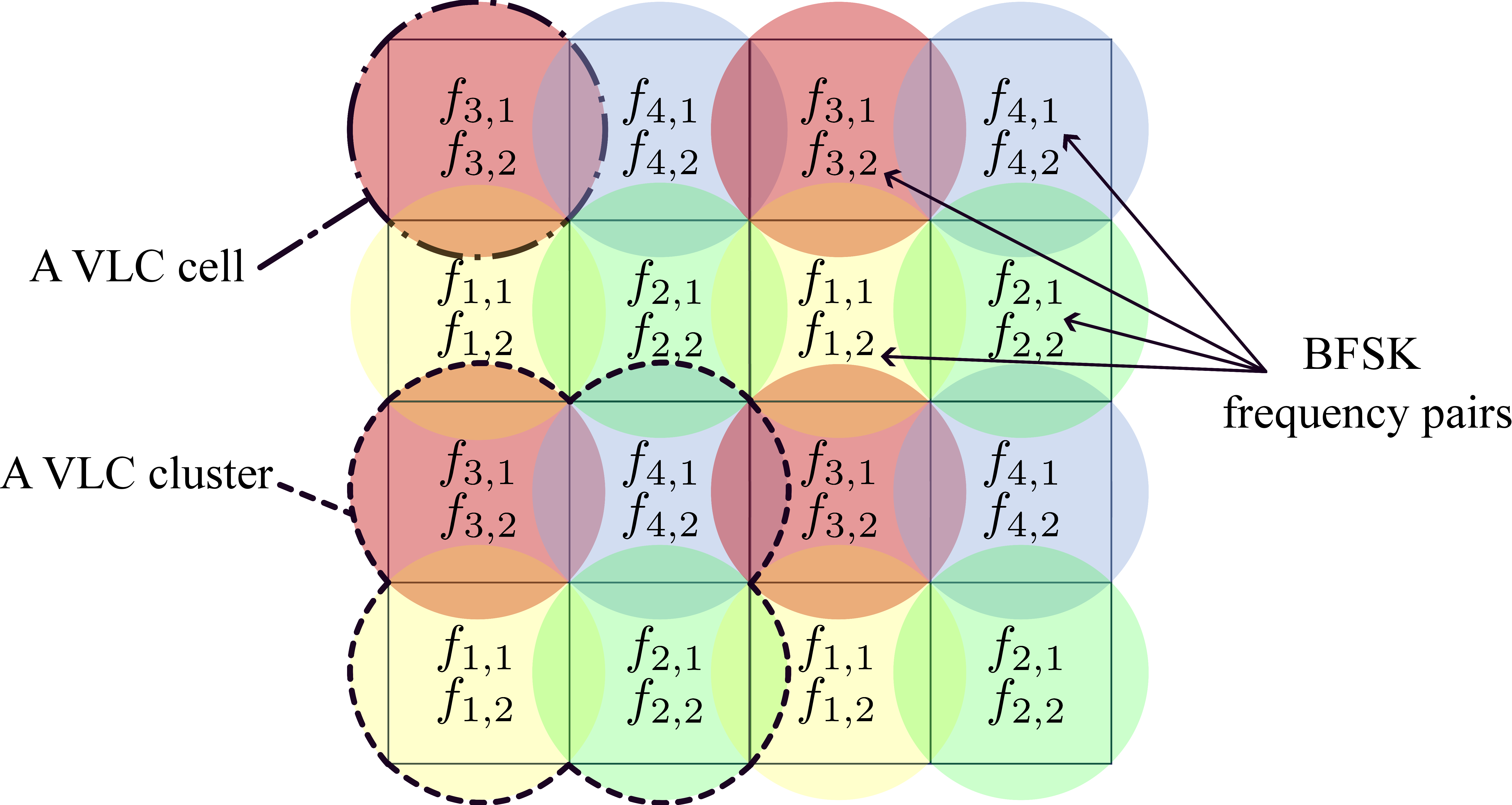}
\caption{An example of an FDM-based VLC deployment. A VLC cluster consists of four VLC cells denoted by four different colors. Within a cluster, each cell adopts a distinct frequency pair for BFSK-based transmission of its VLC cell ID, thereby avoiding interference between adjacent cells.}
\label{fig:cluster_ground}
\vspace{-10pt}
\end{figure}
In the proposed system, capturing VLC data from multiple LED APs concurrently necessitates a multiplexing scheme to distinguish their transmissions.
One option is time-division multiplexing (TDM), where each LED AP transmits in a specific time slot while remaining idle at other times~\cite{cherntanomwong2015prox}. 
However, TDM-based approaches impose strict synchronization requirements and potentially large positioning delays if the number of LED APs is high. 
To avoid these drawbacks, this paper proposes an FDM-based scheme that allocates a unique BFSK frequency pair to each LED AP for data transmission, as shown in Fig.~\ref{fig:cluster_ground}. 
This design is inspired by the classic four-color theorem in graph theory~\cite{robertson1997four,tse2005fundamentals}, which implies that a sufficient number of frequency pairs can be allocated such that adjacent VLC cells do not share the same pair to avoid strong co-channel interference. 
In particular, the proposed cell architecture differs from a generic multi-cell VLC deployment since it is designed for proximity reporting rather than for high-throughput data delivery. Each local VLC cell acts as a frequency-identified proximity region, and the FDM structure is chosen so that several overlapping cells can illuminate the BD simultaneously while still producing separable signatures at the reader.

Fig.~\ref{fig:cluster_ground} presents an example, where each cluster of four LED APs adopts four distinct BFSK frequency pairs. By examining which frequency components appear in the backscattered signal coming from the BD, the RF reader can determine which LED APs illuminate the current position of the BD. 
Therefore, proximity information can be received and forwarded by the BD without significant delay or tight synchronization requirements at the device side.
For example, let $C$ denote the number of locally adjacent or overlapping VLC cells that may illuminate the BD, and let $T_{\textrm{pkt}}$ denote the packet duration of one LED message. Under TDM, the worst-case observation delay for collecting the local LED reports scales as $D_{\textrm{TDM}}^{\textrm{worst}} = C T_{\textrm{pkt}}$, because the relevant LED APs transmit sequentially in different slots. On the other hand, under the proposed concurrent FDM signaling, the corresponding delay remains on the order of $D_{\textrm{FDM}}^{\textrm{worst}} \approx T_{\textrm{pkt}}$, provided that the reader can separate the active frequency branches. Therefore, the local proximity-update latency of FDM is largely insensitive to the size of the local cell cluster, whereas the latency of TDM grows linearly with $C$.

The required number of distinct BFSK frequency pairs is determined by the frequency-reuse patter of the cluster of VLC cells that is used, rather than the total number of luminaires. Under the assumption adopted in this paper, the number of locally distinct pairs is upper-bounded by the chromatic number of the color graph that is built, which is at most four in the idealized case. Consequently, frequency reuse can be applied across spatially separated zones, whereas TDM continues to accumulate slot delay as the number of locally scheduled cells increases.
For large-scale deployment, a practical implementation is to partition the facility into repeated VLC clusters or zones, applying spatial frequency reuse across these non-overlapping zones. The reader can combine the frequency-reuse pattern with a zone-aware illumination map, such that the same frequency pair can then be interpreted unambiguously once the coarse region of the BD is known.
\subsection{Particle Filter-Based Tracking Algorithm Design}
\label{subsec:pf_tracking}
This subsection presents a PF-based tracking framework for the fusion of reader-decoded LED IDs and RSS measurements of backscatter signals. 
The objective is to iteratively estimate the position of the BD under uncertain motion dynamics and imperfect observations.
The PF principle follows standard sequential Monte Carlo methodology; however, the measurement model and the algorithm are tailored to the proposed joint VLC-BC tracking system.
The state-space model and measurement model
are introduced in the following paragraphs.
\subsubsection{State-Space Model}
The 2D motion of the BD in discrete time steps indexed by $k$ is considered.
Let \mbox{$\bm{x}[k] = \bigl[p_{x}[k],\;p_{y}[k],\;v_{x}[k],\;v_{y}[k]\bigr]^{\textrm{T}}$}
indicate the position of the BD $(p_{x}, p_{y})$ and its velocity $(v_{x}, v_{y})$ at time instance~$k$. Then, a standard kinematic evolution model~\cite{bar2004estimation} in discrete time is implemented as follows:
\begin{equation}
\label{eqn:state_evo}
\bm{x}[k+1] = \bm{F}\,\bm{x}[k] + \bm{G}\,\bm{w}[k],
\end{equation}
where $\bm{F}$ and $\bm{G}$ are given by
\begin{equation}
\label{eqn:state_matrix}
\bm{F} = \begin{bmatrix}
1 & 0 & T_{\textrm{s}} & 0 \\
0 & 1 & 0 & T_{\textrm{s}} \\
0 & 0 & 1 & 0 \\
0 & 0 & 0 & 1
\end{bmatrix},\quad
\bm{G} = \begin{bmatrix}
\tfrac{1}{2}T_{\textrm{s}}^{2} & 0\\
0 & \tfrac{1}{2}T_{\textrm{s}}^{2}\\
T_{\textrm{s}} & 0\\
0 & T_{\textrm{s}}
\end{bmatrix},
\end{equation}
respectively. The term $\bm{w}[k]\sim\mathcal{N}(\bm{0},\bm{Q}_{w})$ is zero-mean white Gaussian noise with $\bm{Q}_{w}=\sigma_{w}^2\bm{I}$ capturing random accelerations, where $\bm{I}$ denotes an identity matrix. Moreover, $T_{\textrm{s}}$ is the sampling interval. 
In~\eqref{eqn:state_matrix}, the state transition matrix~$\bm{F}$ represents the linearized 2D kinematic model. The model assumes that the state of the BD is perturbed by random acceleration during motion, represented by the white-noise term $\bm{w}[k]$. The model has been widely used in kinematic tracking systems~\cite{bar2004estimation} and is known as the white-noise acceleration model. 
While this model offers low computational complexity and stable behavior for smooth motion, it can be affected by abrupt maneuvers such as sharp turns. In those cases, the predicted particle propagation model may temporarily mismatch the actual motion of the BD, which increases the positioning error until the measurement update pulls the particle cloud back toward the true trajectory. 

\subsubsection{Measurement Model}
At each time instance $k$, the measurement vector can be expressed by
\begin{equation}
\label{eqn:mea_vector}
\bm{z}[k] = \Bigl\{ \mathcal{I}[k],
\,
\mathcal{R}_{\textrm{dB}}[k]
\Bigr\},
\end{equation}
The measurement vector consists of a set of detected LED IDs, denoted by $\mathcal{I}[k]$, and a measured backscatter RSS value, denoted by $\mathcal{R}_{\textrm{dB}}[k]$. The measured RSS can be expressed as
\begin{equation}
\mathcal{R}_{\textrm{dB}}[k] = R_{\textrm{dB}}^\textrm{BC}\bigl(\bm{x}[k]\bigr) + v[k],
\end{equation}
with $v[k]\sim \mathcal{N}(0,\sigma_{v}^{2})$ denoting the measurement noise. The function $R_{\textrm{dB}}^\textrm{BC}(\cdot)$ is the RSS model of the backscatter signal described in~\eqref{eqn:bc_rss} in logarithmic scale.
The set $\mathcal{I}[k]$ includes all LED APs whose coverage regions contain the position of the BD; if no LED AP illuminates the BD, the set is empty. 
We adopt a low-complexity match-or-penalty rule for comparing the measured and predicted LED sets, namely
\begin{equation}
\label{eqn:likelihood_compare}
p\!\left(\mathcal I^{\textrm{mea}}\mid \mathcal I^{\textrm{pred}}\right)
=
\begin{cases}
1, & \mathcal I^{\textrm{mea}} \cap \mathcal I^{\textrm{pred}} \neq \emptyset,\\
0, & \mathcal I^{\textrm{mea}} \cap \mathcal I^{\textrm{pred}} = \emptyset, \end{cases}
\end{equation}
where $\mathcal{I}^{\textrm{mea}}$ and $\mathcal{I}^{\textrm{pred}}$ represent the measured and predicted LED IDs, respectively. 
Moreover, $\mathcal I^{\textrm{pred}}$ is obtained by checking the IDs of the LED APs that fall inside the FoV of the BD-equipped photodetector at the position of $\bm{x}[k]$, using a known VLC deployment and LED illumination map in the space of interest.
This choice treats the LED-ID observation as a coarse geometric gate, leading to a joint measurement likelihood function that attains the form
\begin{equation} 
\label{eqn:likelihood}
p\bigl(\bm{z}\mid \bm{x}\bigr)
\;\propto\;
p\Bigl(\mathcal{I}^{\textrm{mea}}\mid \mathcal{I}^{\textrm{pred}}\Bigr) \\
\;\times\;
\exp\!\Bigl(\!-\tfrac{\bigl(\mathcal{R}^{\textrm{mea}}_{\textrm{dB}} - \mathcal{R}^{\textrm{pred}}_{\textrm{dB}}\bigr)^2}{2\,R_{v}}\Bigr).
\end{equation}
Similarly, $\mathcal{R}^{\textrm{mea}}_{\textrm{dB}}$ and $\mathcal{R}^{\textrm{pred}}_{\textrm{dB}}$ denote the measured and predicted RSS of the backscatter signal, respectively.
A Gaussian likelihood is applied for the RSS term with noise variance $R_{v}$.
The above binary LED-ID rule is aligned with the low-complexity computation objective at the reader and treats the no-coverage case as compatible when both sets are empty. However, it does not explicitly distinguish between missed detections, false alarms, and partial set mismatches.

\begin{algorithm}
\small
\caption{PF Tracking Algorithm Pseudo-code}
\label{alg:particle_filter}
\KwIn{
measurements $\bm z[k]=\bigl(\mathcal{I}^{\textrm{mea}},\,\mathcal{R}^{\textrm{mea}}_{\textrm{dB}}\bigr)$,
process noise covariance $\bm{Q}_w$, 
measurement noise variance $R_v$
}
\KwOut{State estimates $\hat{\bm{x}}[k]$}

\textbf{Initialization:}\\
Generate $N_{\textrm{p}}$ particles $\{\bm{x}_{i}[0]\}_{i=1}^{N_{\textrm{p}}}~\sim \mathcal{N}(0, \bm{Q}_0)$\\
Assign equal weights $w_{i}[0] = 1/{N_{\textrm{p}}}$ to all particles\\[3pt]

  \textbf{Prediction:}\\
  \For(\tcp*[h]{State evolution}){\text{each particle} $i = 1$ to $N_{\textrm{p}}$}{
    Sample process noise $\bm{w}_{i}[k] \sim \mathcal{N}\bigl(\bm{0},\, \bm{Q}_{w}\bigr)$\\
    Predict particle state using~\eqref{eqn:state_evo}
  }

  \textbf{Measurement Update:}\\
  \For(\tcp*[h]{Weight computation}){\text{each particle} $i = 1$ to $N_{\textrm{p}}$}{
    Compute predictions:
    $
    \bigl(\mathcal{I}^{\textrm{pred}},\,
          \mathcal{R}^{\textrm{pred}}_{\textrm{dB}}\bigr)
    $
    using $\bm{x}_{i}[k]$\\
    Compare with observations: 
    $\bigl(\mathcal{I}^{\textrm{mea}},\,\mathcal{R}^{\textrm{mea}}_{\textrm{dB}}\bigr)$\\
    Evaluate the unnormalized weight:\\
    $
    w_{i}[k] 
    \;=\;
    w_i[k-1]
    \;\times\;
    p\Bigl(\mathcal{I}^{\textrm{mea}}\mid \mathcal{I}^{\textrm{pred}}\Bigr)
    \;\times\;
    \exp\!\Bigl(\!-\tfrac{\bigl(\mathcal{R}^{\textrm{mea}}_{\textrm{dB}} - \mathcal{R}^{\textrm{pred}}_{\textrm{dB}}\bigr)^2}{2\,R_{v}}\Bigr)
    $
    \\
  }
  If $\sum_{j=1}^{N_{\textrm{p}}} w_j[k] > 0$, normalize the weights:
  $
  w_{i}[k] \;\leftarrow\;
  w_{i}[k]\,/\,\sum_{j=1}^{N_{\textrm{p}}} w_{j}[k]
  $;
  otherwise, assign equal weights $w_i[k]=1/N_{\textrm{p}}$\\
  
  \textbf{Resampling:}\\
  Draw $N_{\textrm{p}}$ samples from $\{\bm{x}_{i}[k]\}_{i=1}^{N_{\textrm{p}}}$ according to the normalized weights and replace the old samples\\
  Assign equal weights $w_i[k]=1/N_{\textrm{p}}$ to the resampled particles\\

  \textbf{State Estimate:}\\
  Compute weighted average of particles as the estimated state using~\eqref{eqn:compute_estimate}
\end{algorithm}
\subsubsection{Sequential Monte Carlo Method}
A particle-filter-based approach~\cite{bar2004estimation} is adopted to combine the measurements of LED IDs and the RSS of the backscatter signal for recursively estimating the location of the BD over time.
The PF is specialized to the proposed VLC-BC system through its measurement construction, where overlap-induced LED-ID sets act as discrete geometric constraints and the BC RSS provides continuous refinement of the state estimate.
With the aid of the state-space model~\eqref{eqn:state_evo} and measurement model~\eqref{eqn:mea_vector}, positioning and tracking of the BD are defined as a state estimation problem.
The PF employs the sequential Monte Carlo method~\cite{bar2004estimation}, which approximates a posteriori probability distribution of the state of the BD at each time instance $k$ as
\begin{equation}
    p(\bm{x}[k]|\bm{z}[1\colon k]) \propto p(\bm{z}[k]|\bm{x}[k])\,p(\bm{x}[k]|\bm{z}[1\colon k-1]), 
\end{equation}
where $p(\bm{x}[k]|\bm{z}[1\colon k-1])$ is the prior state and $p(\bm{z}[k]|\bm{x}[k])$ is the likelihood function in~\eqref{eqn:likelihood}. The PF tracks the BD by sequentially updating the state $\bm{x}[k]$ using the measurements $\bm{z}[1\colon k]$.
The pseudo-code in Algorithm~\ref{alg:particle_filter} describes the main loop, including prediction, measurement update, resampling, and state estimation. 
The initial state $\bm{x}[0]\sim\mathcal{N}(0,\bm{Q}_0)$ is represented by $N_\textrm{p}$ particles distributed in the space, where all particles are assigned equal normalized weights $w[0]$.
During the prediction step, the state of each particle is propagated according to~\eqref{eqn:state_evo}. The measurement update then adjusts the weight of each particle based on how well the prediction agrees with measurements, expressed by
\begin{equation}
    w_i[k] = w_i[k-1] \, p(\bm{z}[k]|\bm{x}_i[k]).
\end{equation}
In particular, the set of predicted LED IDs $\mathcal{I}^{\textrm{pred}}$ is obtained by checking which LED APs cover the location of each particle.
The observed LED ID set $\mathcal{I}^{\textrm{mea}}$ is compared with the predicted set $\mathcal{I}^{\textrm{pred}}$ through~\eqref{eqn:likelihood_compare}. 
Hence, in the current design, the LED-ID term acts as a coarse feasibility constraint on the particle location; that is, particles that are non-compatible with the measured illuminated-cell set are rejected, whereas particles that remain compatible are further differentiated by the RSS term.
A more refined likelihood model can be implemented to distinguish between mild and severe LED-ID set mismatches, although it would add computational complexity.
The RSS measurement of the backscatter signal refines this process by penalizing large deviations between the measured RSS $\mathcal{R}^{\textrm{mea}}_\textrm{dB}$ and predicted RSS $\mathcal{R}^{\textrm{pred}}_\textrm{dB}$, thereby incorporating the backscatter channel condition into the likelihood calculation. 
A resampling step is performed to avoid the degeneracy problem discussed in~\cite{bar2004estimation}. Particles with low weights are replaced by particles with higher weights to maintain particle diversity and prevent weight collapse. 
Finally, the position of the BD is estimated as the posterior mean of the state over all particles using
\begin{equation}
\label{eqn:compute_estimate}
\hat{\bm{x}}[k] = \sum_{i=1}^{N_{\textrm{p}}} w_{i}[k]\;\bm{x}_{i}[k].
\end{equation}

The above BD-position tracking process is repeated recursively to provide continuous state estimation.
This combination of proximity-based LED-ID matching and RSS refinement provides a low-complexity solution for positioning and tracking. It ensures that proximity information from multiple LED APs imposes coarse constraints on the location of the BD, while the RSS observation provides a continuous measurement that refines the estimate.
\subsubsection{Computational Complexity Analysis}
Let $N_\textrm{p}$ be the number of particles, $L$ the number of LED APs or VLC cells, and $d=4$ the dimension of the state vector $\bm{x}$. 
The prediction step costs $O(N_\textrm{p})$. 
The LED-ID coverage evaluation tests each particle against $L$ VLC cells and therefore costs $O(N_\textrm{p} L)$ in the worst case, which dominates when $L\!\ge\!2$.
Furthermore, the RSS likelihood calculation, weight normalization, resampling, and posterior averaging require $O(N_\textrm{p})$ for each step. 
Summing these contributions gives the worst-case per-step time complexity \mbox{$\mathcal{C} = O(N_\textrm{p} L)$}. 
Moreover, the working memory scales linearly with the number of particles, thereby yielding a memory complexity of $O(N_\textrm{p} d)$.
This expression is a worst-case bound. In practice, a particle does not need to be checked against all LED APs in a large deployment scenario, because only a small subset of nearby VLC cells can cover a given position. Therefore, the reader can reduce the average complexity by using a precomputed spatial index, occupancy grid, or adjacency list that returns only a local candidate set of LEDs for each particle. If the average candidate count is denoted by $\bar{L}$, with $\bar{L}\ll L$, the effective per-step cost becomes closer to $O(N_\textrm{p}\bar{L})$.
This observation also highlights the importance of the particle count $N_\textrm{p}$ as a design parameter. A larger $N_\textrm{p}$ may improve the approximation accuracy of the posterior distribution, yet increases the computational cost at the reader. 
To quantify this trade-off, Section~\ref{subsec:experiments} includes a sensitivity study over various $N_\textrm{p}$.
\section{Implementation}\label{sec:implementations}
\begin{figure*}
\centering
\includegraphics[width=1.98\columnwidth]{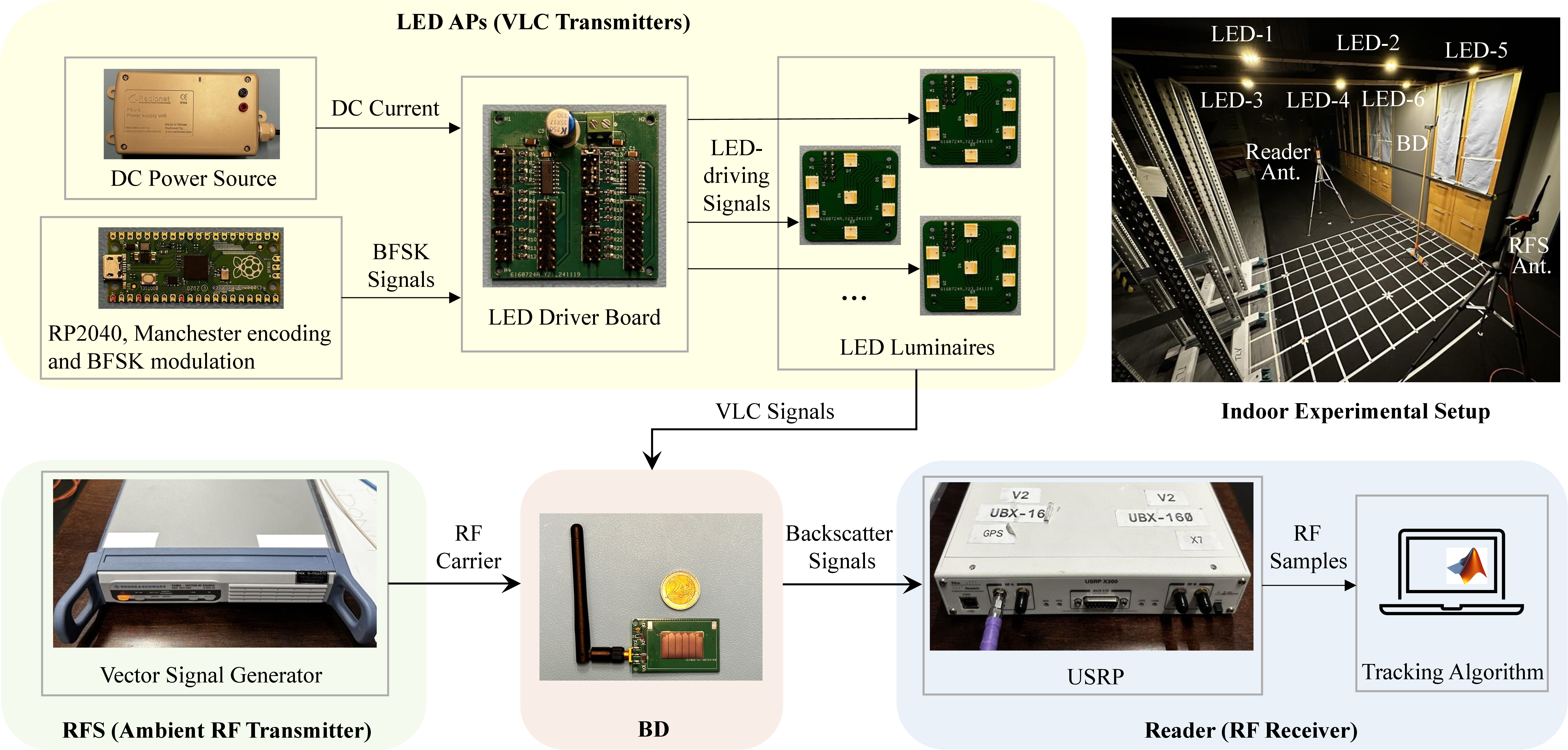}
\caption{Implementation of the proof-of-concept system. 
Six LED luminaires emit VLC signals embedded with their IDs. 
A signal generator emits 2.4~GHz carrier waves. 
The BD receives the VLC signals, converts them into electrical signals, and uses them to modulate the carrier signal.
The modulated signals are backscattered and captured by a USRP, which is connected to a host computer for signal processing and location inference of the BD. The system is evaluated in an indoor space.}
\label{fig:implement_diagram}
\end{figure*}
\begin{table}
\centering
\caption{Hardware Components Used in the Implementation}
\scriptsize
\begin{tabular}{c|c|c}
\hline
\textbf{Parameter}               & \textbf{Key Component} & \textbf{Model}          \\ \hline
\multirow{4}{*}{LED AP}          & DC supply              & Radionet PSU-3          \\ \cline{2-3} 
                                 & MCU (AC supply)        & Raspberry Pi RP2040        \\ \cline{2-3} 
                                 & Driver chip                 & OnSemi CAT4109          \\ \cline{2-3} 
                                 & LED                    & OSRAM GW J9LHS2.4M      \\ \hline
\multirow{4}{*}{BD}              & PV cell                & TDK BCS2717B6           \\ \cline{2-3} 
                                 & Comparator             & TI TLV7031              \\ \cline{2-3} 
                                 & RF switch              & Analog Devices ADG919   \\ \cline{2-3} 
                                 & Antenna                & Omni-directional dipole \\ \hline
\multirow{2}{*}{RFS}             & Signal generator       & R\&S SGT100A            \\ \cline{2-3} 
                                 & Antenna                & Omni-directional dipole \\ \hline
\multirow{3}{*}{Reader}          & SDR                    & NI USRP X300            \\ \cline{2-3} 
                                 & Antenna                & Omni-directional dipole \\ \cline{2-3} 
                                 & Laptop                 & Dell Latitude 7420      \\ \hline
\end{tabular}
    \label{tab:hardware}
    \end{table}
A proof-of-concept system is developed to evaluate the feasibility and performance of the proposed design. Fig.~\ref{fig:implement_diagram} shows the experimental setup, and Table~\ref{tab:hardware} summarizes the electronic components and measurement equipment used. Experiments are performed in an indoor environment with six LED luminaires mounted on the ceiling to provide illumination and proximity information. Each luminaire comprises seven LEDs that transmit the same VLC signal and broadcast the luminaire's unique ID using the FDM method discussed in Section~\ref{subsec:vlc_deployment}.
\subsection{Hardware Design and Configuration}
\label{subsec:hardware}
\begin{figure}
\centering
\includegraphics[width=0.98\columnwidth]{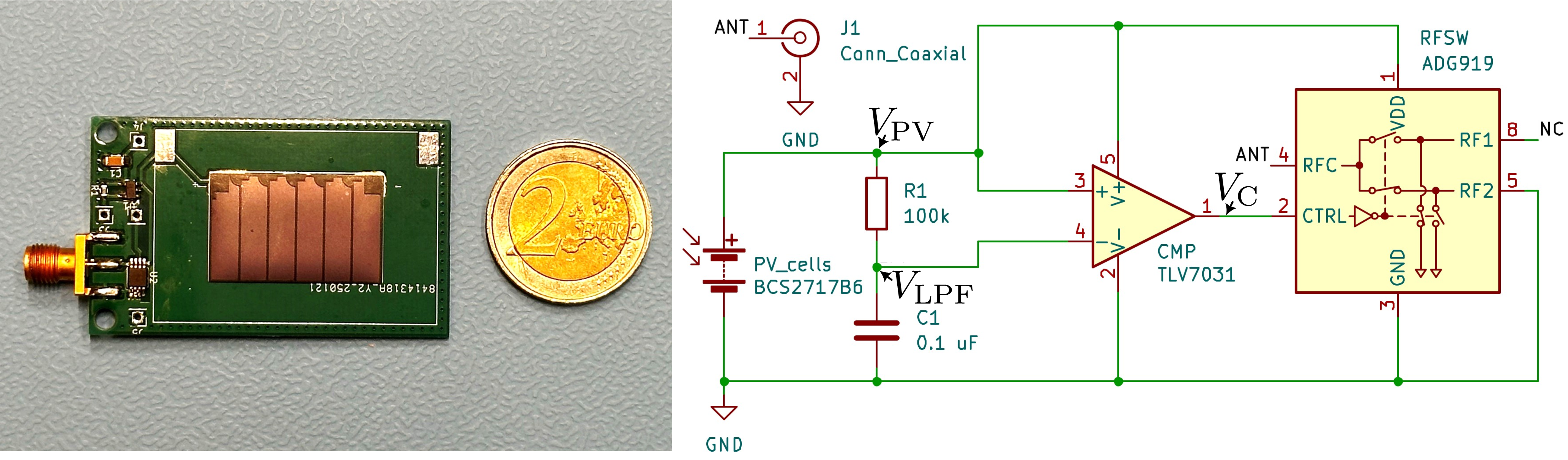}
\caption{Prototype of the BD with its schematic. A 2-euro coin is placed in the picture to show the scale of the BD prototype.}
\label{fig:bd}
\vspace{-10pt}
\end{figure}
A Raspberry Pi RP2040 microcontroller unit (MCU) is used to generate the LED-modulating signals for the luminaires. These signals are generated using Manchester encoding and BFSK modulation. 
The frame structure of the modulating signal consists of a preamble for signal detection and synchronization, followed by a positioning payload that includes an 8-bit LED ID and a dummy bit. 
The preamble is the 7-bit Barker sequence ``1110010,'' which enhances the autocorrelation properties used for detection. 
An LED driver module based on two OnSemi CAT4109 chips is then employed to add a DC bias from a constant power supply to the modulating signals. 
The driver module drives the six LED luminaires simultaneously, with each luminaire emitting a BFSK-modulated signal generated by a dedicated programmable input/output pin of the MCU. 
The BFSK modulation frequencies are chosen to be higher than 2~kHz to eliminate visible flicker in the emitted light. 
Beyond the flicker constraint, the selected frequency pairs are also chosen by considering the practical bandwidth of the end-to-end prototype. More specifically, the frequencies should remain sufficiently separated for filter-bank detection at the reader, while also staying within a modulation range that can be supported reliably by the photovoltaic-comparator front end of the BD.
Referring to the VLC deployment strategy discussed in Section~\ref{subsec:vlc_deployment},
a complete VLC cluster is formed by \mbox{LED-1}, \mbox{LED-2}, \mbox{LED-3}, and \mbox{LED-4}, with each using a distinct pair of BFSK frequencies $\{f_{i,1}, f_{i,2}\}$ for $i\in\{1,2,3,4\}$. Another incomplete VLC cluster includes \mbox{LED-5} and \mbox{LED-6}, which use the same frequency pairs as LED-1 and LED-3, respectively.
The numerical values of the frequencies used are listed in Table~\ref{tab:parameters}.
This design enables continuous broadcasting of the ID assigned to each luminaire via the VLC link. When the BD detects light from one or more luminaires, it reports the corresponding IDs, thereby indicating its proximity to those luminaires.

Fig.~\ref{fig:bd} shows the BD prototype with its circuit diagram. A key point is that the BD operates as an analog light-to-RF converter rather than as a conventional digital VLC receiver aiming at decoding the transmitted sequence of symbols. The PV cell TDK BCS2717B6 harvests light energy from the luminaires and converts the incident optical waveform into an electrical voltage $V_{\textrm{PV}}$. The voltage $V_{\textrm{PV}}$ is fed into the positive input of a comparator Texas Instruments TLV7031 and an RC low-pass filter (LPF) consisting of a 100~k$\Omega$ resistor and a 0.1~$\mu$F capacitor with a 15~Hz cutoff frequency. The LPF output, denoted by $V_{\textrm{LPF}}$, tracks the slowly varying DC component of $V_{\textrm{PV}}$ and is connected to the negative comparator input. When the instantaneous $V_{\textrm{PV}}$ exceeds $V_{\textrm{LPF}}$, the comparator output $V_{\textrm{C}}$ rises to the high binary level. Otherwise, $V_{\textrm{C}}$ falls to the low binary level.
When the BD is illuminated by multiple LED APs simultaneously, the photovoltaic output is a superposition of the corresponding modulated optical components. Denoting by $\mathcal{S}$ the set of illuminating LEDs, the front-end voltage can be interpreted as
$
V_{\textrm{PV}}(t) \approx V_{\textrm{DC}} + \sum_{l\in\mathcal{S}} a_l m_l(t),
$
where $V_{\textrm{DC}}$ is the slowly varying bias term and $a_l m_l(t)$ is the AC data-carrying signal contribution of $\textrm{LED-}l$. The comparator then produces a waveform according to the sign of $V_{\textrm{PV}}(t)-V_{\textrm{LPF}}(t)$. Hence, the BD does not separately demodulate or decode individual VLC packets in overlapping regions. Instead, it forwards a composite binary switching waveform that contains the active frequency components originating from all incident LEDs.

The comparator output controls the Analog Devices ADG919 RF switch, which modulates the incident RF carrier waves captured by the dipole antenna. The termination of the antenna is alternated between two loads $Z_{1}$ and $Z_{2}$. Assuming that the antenna is matched to a characteristic impedance \mbox{$Z_{\textrm{a}}=50~\Omega$}, the complex reflection coefficient~\cite{griffin2009linkbudget} takes the form \mbox{$\Gamma_{i} = \frac{Z_{i}-Z_{\textrm{a}}^{*}}{Z_{i}+Z_{\textrm{a}}}$}, and thus the modulation factor~\cite{griffin2009linkbudget} is expressed by \mbox{$M = \frac{1}{4}\bigl|\Gamma_{1}-\Gamma_{2}\bigr|^{2}$}.
The maximum efficiency \mbox{$M=1$} can be achieved when \mbox{$Z_{1} = \infty$} and \mbox{$Z_{2} = 0$}, corresponding to the switch being either open or shorted at the antenna termination. 

The BD harvests energy from visible light to power its circuitry, enabling energy-neutral operation.
For instance, at an illuminance of 200~lx, the selected PV cell delivers an operating current of 10~\textmu A at 2.6~V, which corresponds to about 26~\textmu W of harvested electrical power. 
For the electronic components of the BD at the same supply voltage, the RF switch draws about 0.10~\textmu A under typical conditions (maximum 1.0~\textmu A), and the comparator draws about 0.315~\textmu A under typical conditions (maximum 0.90~\textmu A). 
These figures give a typical BD power of roughly 1.08~\textmu W and a maximum power of roughly 4.94~\textmu W at 2.6~V. 
Thus, under the stated illumination and operating conditions, the output power from the PV cell exceeds the power consumption of the BD,
supporting energy-neutral operation of the device.
In addition to its low power consumption, the BD has a modest tag-side material cost. Based on representative one-unit pricing of the main prototype components, the cost of the BD is approximately 4.41~EUR excluding assembly and laboratory instrumentation costs. This estimate covers the photovoltaic cell, comparator, RF switch, antenna, small passive components, and a simple PCB/interconnect realization. Therefore, the low-cost description refers specifically to the tag-side hardware of the BD, rather than to the total cost of the full infrastructure.

A Rohde \& Schwarz SGT100A RF signal generator emits an unmodulated sine-wave carrier signal at $f_{\rm c}=2.4$~GHz, serving as the RFS. 
In the proof-of-concept implementation, a signal generator is used to guarantee that the RF carrier power and frequency remain stable and controllable during the experimental measurements. 
In practice, the RF carrier source can be provided by existing wireless access infrastructure, such as Wi-Fi APs and cellular base stations.
When the carrier signal arrives at the antenna, it is mixed with the switch-control signal $V_{\rm CTRL}$ through the switching operation. 
The modulated signal, which carries the LED IDs, is backscattered to the reader. 
The RF reader uses a National Instruments software-defined radio~(SDR) USRP X300, which collects RF samples at a sampling rate $f_{\textrm{s}}=200$~kHz. The samples are conveyed to a host computer, which detects the LED IDs, measures the RSS of the backscatter signals, and executes the tracking algorithm described in Section~\ref{subsec:pf_tracking}. 
The proof-of-concept setup is laboratory-level and is not optimized to minimize the system-level power consumption. The purpose of the current implementation is controlled validation of the proposed VLC-BC tracking principle, rather than minimizing infrastructure-side energy consumption.
\subsection{Signal Processing at Reader}
\label{subsec:signal_processing}
\begin{figure}
\centering
\includegraphics[width=0.85\columnwidth]{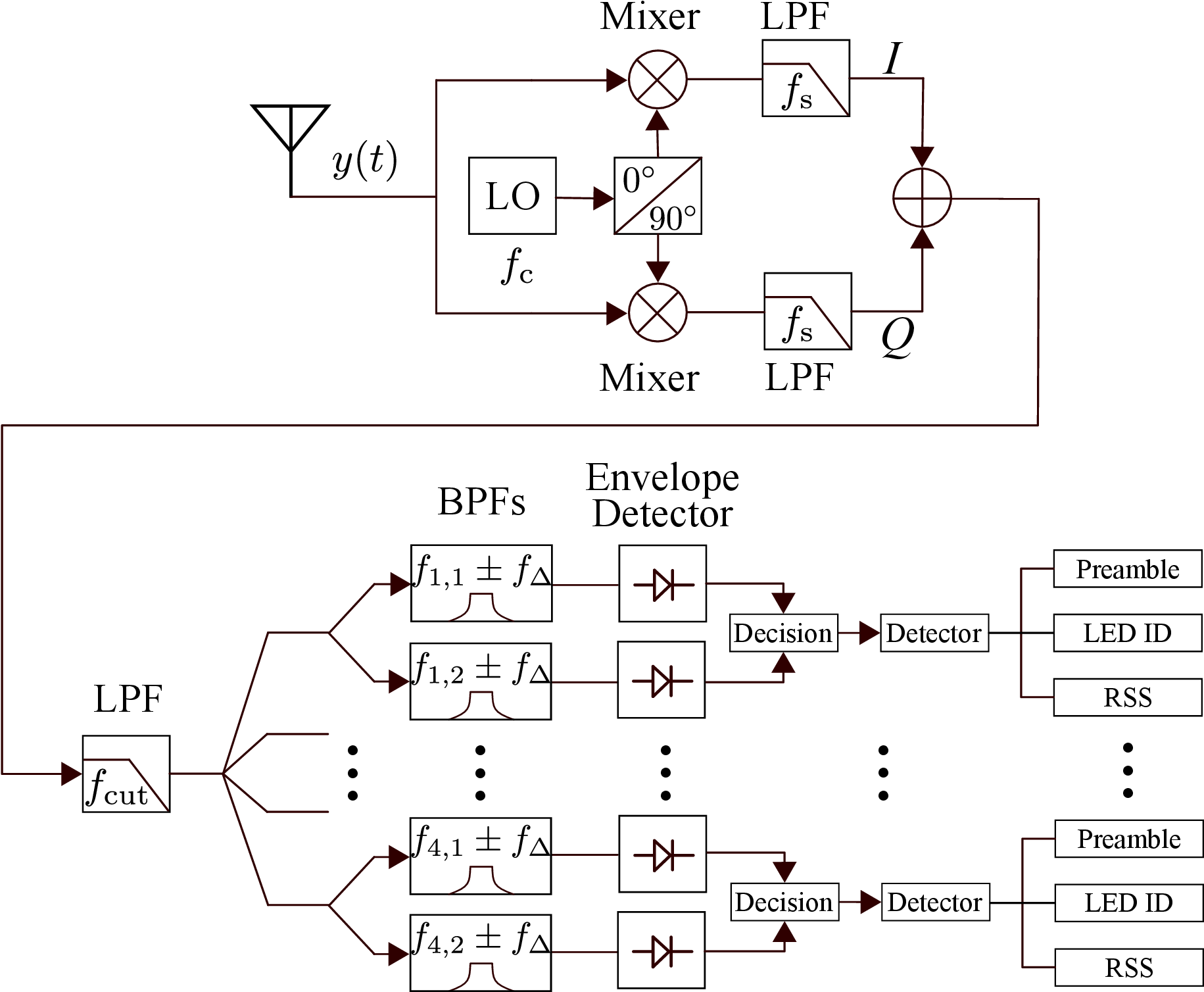}
\caption{Signal processing flow of the received backscatter signals at the reader.} 
\label{fig:rx_schematic}
\vspace{-15pt}
\end{figure}
The signal processing flow at the reader is visualized in Fig.~\ref{fig:rx_schematic}.
The USRP-received RF samples centered at the carrier frequency $f_{\textrm{c}}$ are first down-converted and then low-pass filtered, resulting in a stream of in-phase and quadrature (I/Q) samples at the reader side. These samples are quantized into complex values and transferred to a laptop running MATLAB R2024b for demodulation. A non-coherent FSK demodulator is designed to process the received samples~\cite{goldsmith2005}.
After filtering out the high-frequency noise using an LPF with an approximate cut-off frequency \mbox{$f_\textrm{cut} = 20$~kHz} at 3~dB, a filter bank with four pairs of band-pass filters (BPFs) is applied to the incoming signal samples. Each pair of filters has a 10~dB bandwidth of approximately \mbox{$f_\Delta = 500$~Hz} and is allocated to one specific BFSK frequency pair introduced in Section~\ref{subsec:simulations}.  
This parallel structure is the key mechanism that enables recovery of multiple
LED IDs when the BD is located in an overlap region of several VLC cells.
The outputs of the two BPFs in each pair are envelope-detected and compared to determine which has the greater magnitude, thereby indicating the bit associated with that time instant. The bit sequence produced by this comparison is then examined to detect a preamble, after which the LED AP ID is extracted.
More specifically, each BPF pair isolates one candidate BFSK branch from the composite backscattered signal. The outputs of the two BPFs in that branch are envelope-detected at the reader and compared to determine the instantaneous bit value. The resulting bit stream is then processed independently for preamble detection and synchronization by searching for the Barker sequence, after which the corresponding LED ID is extracted. 
Since this procedure is executed for all assigned frequency pairs in parallel, the reader can recover several LED IDs within the same observation interval whenever several VLC cells illuminate the BD simultaneously.
The extracted ID is combined with the measured RSS to form the measurement vector in~\eqref{eqn:mea_vector}. These quantities are finally passed to the tracking algorithm described in Section~\ref{subsec:pf_tracking}, where they serve as the measurement inputs for tracking the position of the BD.

If the system is extended to a multi-BD scenario, two reader-centric separation mechanisms are compatible with the present architecture. 
Code-division multiple access (CDMA) can be incorporated in the reader, by assigning each BD a tag-specific spreading sequence that gates the switch-control waveform before backscatter modulation. 
Successive interference cancellation (SIC) can be incorporated by allowing overlapping tag transmissions and decoding the strongest response first, reconstructing its contribution, subtracting it from the received signal, and then iterating on the residual signal.

This setup provides a practical implementation of the joint VLC-BC system for tracking, enabling the low-complexity and energy-neutral BD to receive and forward proximity information without an external power supply or an active RF synthesizer.
A long-term reliability study should evaluate the durability of the assembled device under prolonged operation and include a life-cycle analysis.
The setup is evaluated in an indoor environment. 
The experimental results are presented and discussed in Section~\ref{sec:sim_exp_results}. 
\section{Simulation and Experimental Results}
\label{sec:sim_exp_results}
We conduct a series of simulations and experiments to evaluate the performance of the proposed indoor tracking system. This section presents the methodology and findings of both the simulations and the experimental evaluation.
\subsection{Simulations}\label{subsec:simulations}
    \begin{table}[!t]
    \centering
    \caption{Simulation and Experimental Parameters}
    \resizebox{\columnwidth}{!}{%
    \begin{tabular}{l|l|l}
        \hline
        \textbf{Parameter}          & \textbf{Value}    & \textbf{Description}\\ \hline
        $W \times L$                & $2.5 \times 2.5$        & Size of the target area [m$^2$]        \\
        $h_{\text{LED}}, h_{\text{BD}}$ & 1.90, 1.57     & Height of LED APs and BD [m]\\
        $h_{\text{RFS}}, h_{\text{Reader}}$ & 1.5, 1.5      & Height of RFS and Reader [m]\\
        $(x,y)_{\textrm{RFS}}$      & (1.2, -0.5)          & 2D position of RFS [m]\\
        $(x,y)_{\textrm{Reader}}$   & (1.2, 2.0)           & 2D position of reader [m]\\
        $(x,y)_{\textrm{LED}}$      & \makecell{(0.4, 0.4) (1.2, 0.4)\\ (0.4, 1.2) (1.2, 1.2)\\ (2.0, 0.4) (2.0, 1.2)}& 2D positions of LED APs 1-6 [m]\\
        \hline
        $A_{\text{PD}}$             & 0.027$\times$0.017      & Active area of the BD-equipped PD (PV cell) [m$^2$]\\
        $\Phi_{\text{max}}$         & 60    & LED radiation semi-angle at half power [deg]\\
        $\Psi$                      & 60    & FoV semi-angle of the PD’s light acceptance cone [deg]\\
        $\eta_{\textrm{E-O}}$       & 2.1   & LED power electric-optical conversion factor [W/A]\\
        $\eta_{\textrm{O-E}}$       & 0.5   & Responsivity of the BD-equipped PD   [A/W]\\
        $B$                         & 50        & System operational bandwidth [kHz]\\
        $\varepsilon$               & 0.75      & Fill factor\\
        $I_{\text{Bias}}$           & 0.75      & LED driving current [A]\\
        $I_{0}$                     & $10^{-9}$ & Dark saturation current [A]\\
        $V_{\textrm{t}}$             & $25\times 10^{-3}$      & Thermal voltage [V]\\
        $(f_{i,1},f_{i,2})$         &\makecell{(8.254, 9.004)\\(10.074, 10.990)\\(11.918, 13.002)\\(13.742, 14.992)}                    & Frequency pairs of light-modulating signal [kHz]\\
        \hline
        $P_{\text{c}}$              & 20                      & RFS carrier power [dBm]\\
        $f_{\text{c}}$              & 2.4                     & RFS carrier frequency [GHz]\\
        $G_{\text{T}}$, $G_{\text{R}}$,  $G_{\text{BD}}$     & 3, 3, 1.5  & Antenna gain of RFS, reader, and BD [dBi]\\
        $\chi_{\textrm{f}}, \chi_{\textrm{b}}$           & 0.5, 0.5  & Polarization mismatch\\
        $M$                         & 0.5               & Modulation factor\\
        $\Theta$                    & 1 (0~dB)           & On-object penalty factor\\
        $R$                         & 1~kbit/s           & System data rate\\
        $N_0$                       & -174               & Noise power spectral density [dBm/Hz]\\
        $(\sigma_{\text{LoS}},\sigma_{\text{NLoS}})$ & (3, 8.03) & Shadowing factors in the 3GPP model [dB]\\
        \hline
        $N$                         & 5000                  & Number of particles\\
        $T_\textrm{s}$              & 0.2                   & Sampling interval [s]\\
        $\sigma_{v}$                & 5                     & Standard deviation of RSS measurement noise [dB]\\
        $\sigma_{w}$                & 1                     & Standard deviation of process noise\\
    \hline
    \end{tabular}}
    \label{tab:parameters}
    \vspace{-10pt}
    \end{table}
\begin{figure*}
  \centering
  \hspace{0.1\columnwidth}
  \subfloat[]
  {\includegraphics[width=0.5\columnwidth]{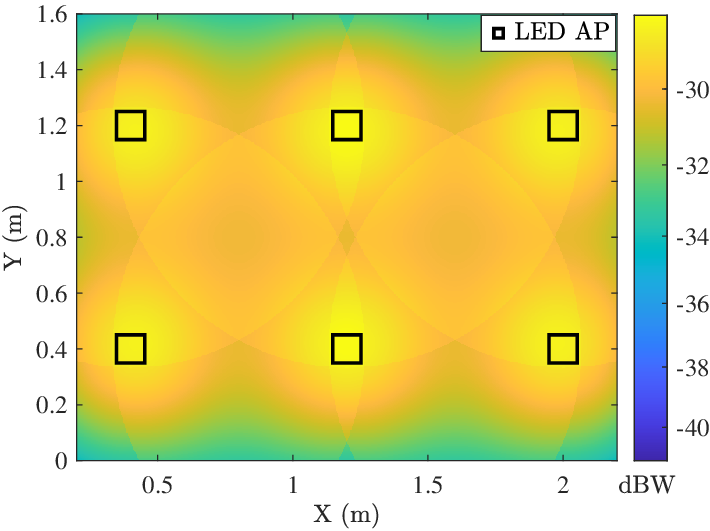}
   \label{fig:sim_vlc_power_h_140}}
  \hspace{0.1\columnwidth}
\subfloat[]
  {\includegraphics[width=0.5\columnwidth]{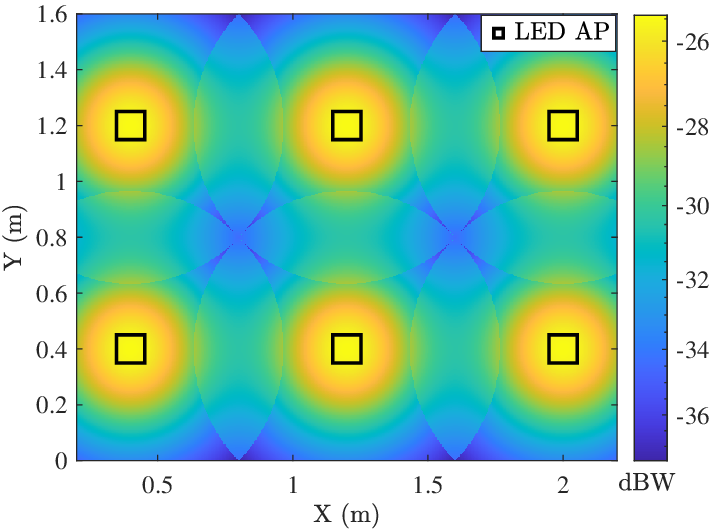}
   \label{fig:sim_vlc_power_h_157}}
  \hspace{0.1\columnwidth}
  \subfloat[]
  {\includegraphics[width=0.5\columnwidth]{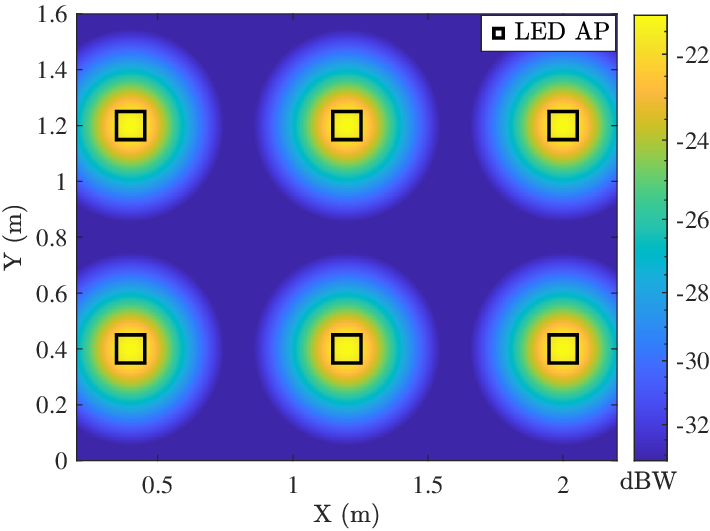}
   \label{fig:sim_vlc_power_h_170}}
  \hspace{0.1\columnwidth}
  \caption{Heatmap of the simulated VLC signal power at the BD over the space of interest. 
  Colors range from dark blue to bright yellow, corresponding to low through high received power levels of the VLC signal.
  Different vertical distances between LED APs and the BD are demonstrated, yielding different sizes of VLC cells:
  (a) $h_{\textrm{BD}} = 1.40$~m ($r_\textrm{cell}=0.87$~m), (b) $h_{\textrm{BD}} = 1.57$~m ($r_\textrm{cell}=0.57$~m), (c) $h_{\textrm{BD}} = 1.70$~m ($r_\textrm{cell}=0.35$~m).
  }
  \label{fig:sim_vl_power}
\end{figure*}
A rectangular space of dimensions $2.5\times 2.5$~m$^2$ is considered, with six LED APs mounted on the ceiling at a height of $h_{\textrm{LED}} = 1.90$~m.  
The LED APs are arranged in a grid with spacing $d_{\textrm{LED}} = 0.8$~m to ensure dense illumination and coverage. All LED APs emit visible light at an optical power of 2.1~W while continuously transmitting packets via VLC using the proposed FDM method. 
Each packet consists of a unique 8-bit ID of the LED AP, preceded by a 7-bit Barker preamble for synchronization. 
An unmodulated sine-wave RF carrier at $f_{\textrm{c}}=2.4$~GHz and a power of 20~dBm is generated by an omnidirectional dipole antenna positioned at \mbox{$(x,y,z)=(1.2, -0.5, 1.5)$~m}. The RF reader, equipped with an omnidirectional dipole antenna and positioned at \mbox{$(x,y,z)=(1.2, 2.0, 1.5)$~m}, samples the received signal at 200~kSa/s. 
Received samples are then processed by the non-coherent FSK demodulator discussed in Section~\ref{subsec:signal_processing} to extract LED IDs and measure the RSS of the backscatter signals. 
The decoded LED IDs and measured RSS values are then provided to the tracking algorithm discussed in Section~\ref{subsec:pf_tracking} to estimate the position of the BD.
The main parameters used in the simulations to characterize the VLC link and BC link are summarized in Table~\ref{tab:parameters}.

Since the size of the VLC cell affects the coverage where the BD can capture light and report its proximity, the aggregated VLC signal power received at the BD is simulated in the region of interest for different cell sizes.
Different vertical distances between LED APs and the BD are considered to vary the size of VLC cells. As examples, three vertical positions of the BD at 1.40, 1.57, and 1.70~m from the floor level are tested, while the LED-AP heights remain fixed. As discussed in Section~\ref{subsec:vlc_cell_model}, a lower BD height yields larger VLC-cell coverage, whereas a higher height results in a smaller VLC cell size. Fig.~\ref{fig:sim_vl_power} illustrates the resulting heatmaps of the VLC signal power arriving at the BD. Colors range from dark blue to bright yellow, corresponding to low through high power levels. 
Each LED AP is observed to illuminate an approximately circular area, referred to as a VLC cell. The overlapping areas of these circles indicate regions where the BD can simultaneously receive multiple VLC signals through the aforementioned FDM method. 
When the vertical position of the BD increases, the received VLC signal power from the nearest LED AP grows as well because of the shorter VLC link distance, whereas the coverage region shrinks and the overlapping zones between LED cells decrease.

\begin{figure}
\vspace{-20pt}
  \centering
  \subfloat[]
  {\includegraphics[width=0.48\columnwidth]{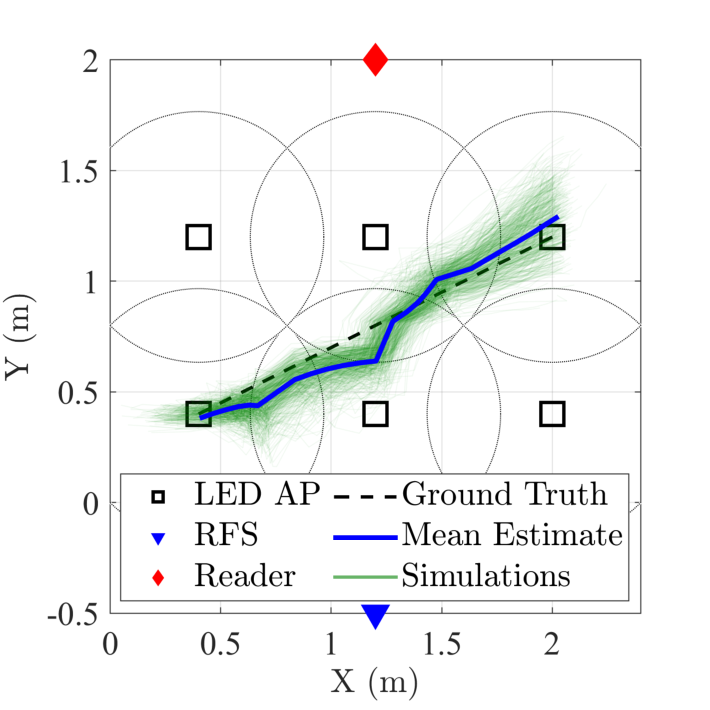}
   \label{fig:sim_track_path_1}}
  \hfill
  \subfloat[]
  {\includegraphics[width=0.48\columnwidth]{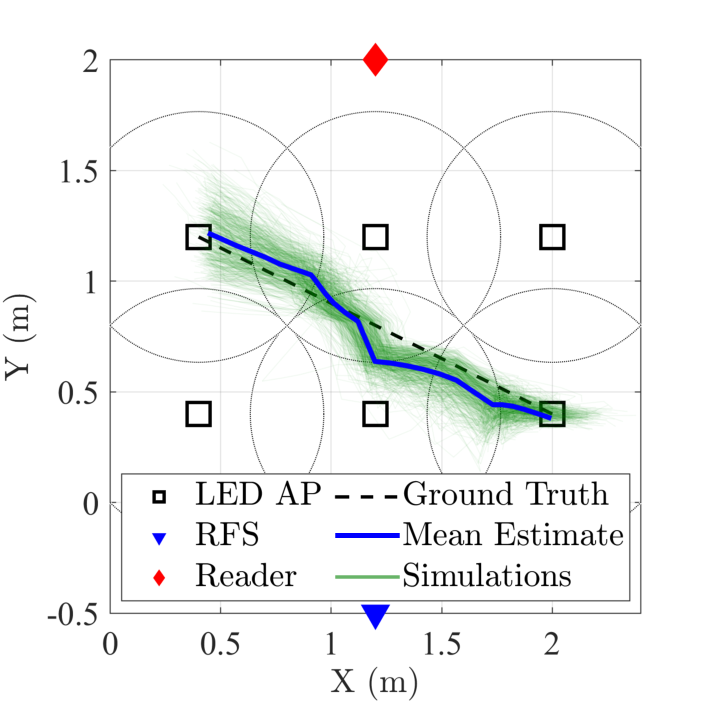}
   \label{fig:sim_track_path_2}} 
  \hfill
  \subfloat[]
  {\includegraphics[width=0.48\columnwidth]{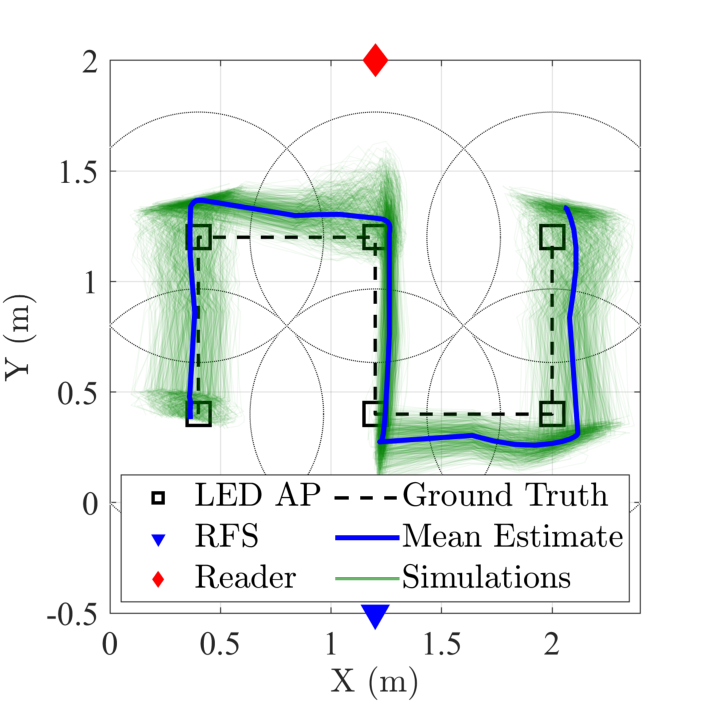}
   \label{fig:sim_track_path_3}}
  \hfill
  \subfloat[]
  {\includegraphics[width=0.48\columnwidth]{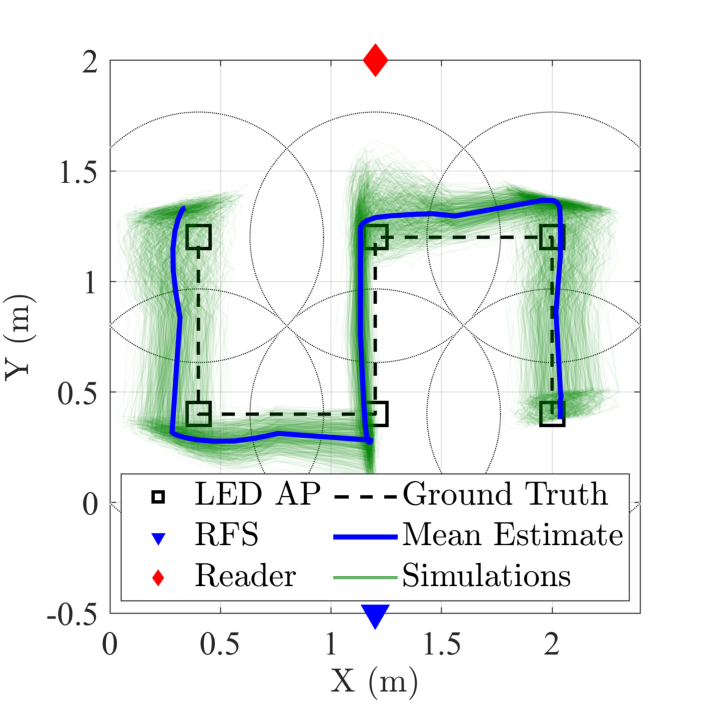}
   \label{fig:sim_track_path_4}} 
  \caption{Monte Carlo tracking results from 500 simulation repetitions for multiple trajectories: (a)-(b) straight-line paths 1 and 2, and (c)-(d) zigzag paths 3 and 4.
  In each plot, the black dashed line is the ground truth of BD movement, the green lines show the individual simulation repetitions, and the blue solid line indicates the mean estimate. Arrows illustrate the directions of BD movement.}
  \label{fig:all_sim_track}
\vspace{-15pt}
\end{figure}
\begin{figure}
  \centering
  \subfloat[]
  {\includegraphics[width=0.48\columnwidth]{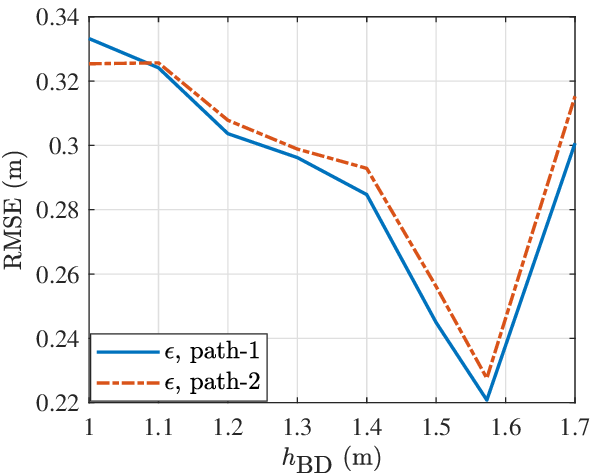}
   \label{fig:sim_rmse_straight}}
  \hfill
  \subfloat[]
  {\includegraphics[width=0.48\columnwidth]{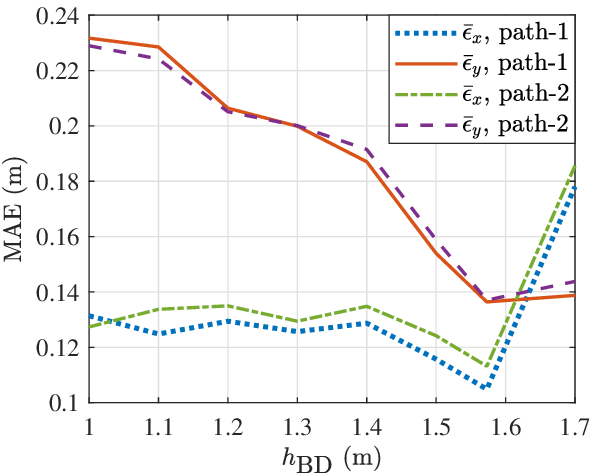}
   \label{fig:sim_mae_straight}} 
  \hfill
  \subfloat[]
  {\includegraphics[width=0.48\columnwidth]{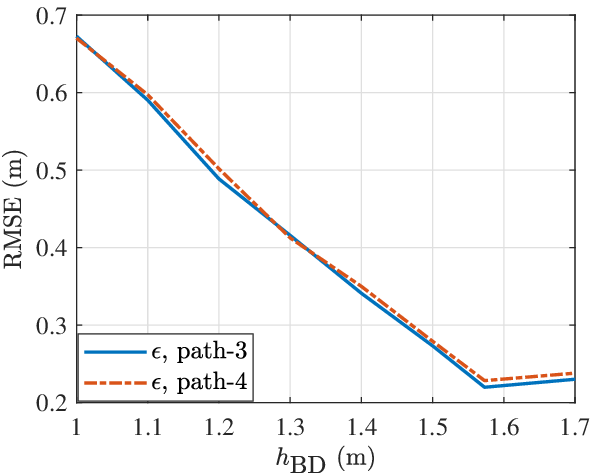}
   \label{fig:sim_rmse_zigzag}}
  \hfill
  \subfloat[]
  {\includegraphics[width=0.48\columnwidth]{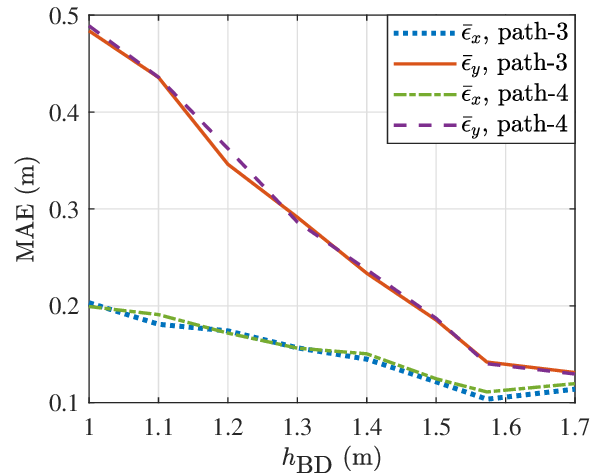}
   \label{fig:sim_mae_zigzag}}
  \caption{Tracking-error evaluation over 500 simulation repetitions for different trajectories and BD heights: (a)-(b) RMSE and MAE of straight-line paths 1 and 2, and (c)-(d) RMSE and MAE of zigzag paths 3 and 4.
  }
  \label{fig:all_sim}
\vspace{-10pt}
\end{figure}
\begin{figure}
\vspace{-5pt}
\centering
\includegraphics[width=0.6\columnwidth]{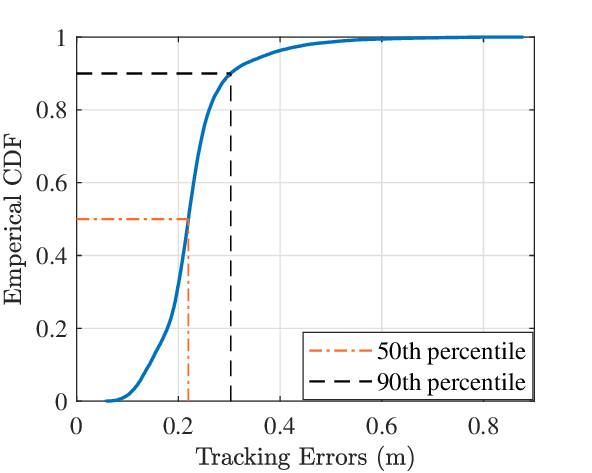}
\caption{Empirical CDF of the positioning errors from simulations across all trajectories.} 
\label{fig:sim_cdf_all}
\vspace{-10pt}
\end{figure}
Monte Carlo simulations are performed in MATLAB R2024b to evaluate the positioning accuracy. The positioning error is measured by root-mean-square error (RMSE) and mean absolute error (MAE) in the \mbox{$x$-direction} and \mbox{$y$-direction}. The RMSE is defined as
\begin{equation}
    \epsilon = \sqrt{\tfrac{1}{K}\sum_{k=1}^{K}\bigl\Vert\bm{p}[k]-\hat{\bm{p}}[k]\bigr\Vert^{2}},
\end{equation}
where $\bm{p}[k]=(p_{x}[k],\,p_{y}[k])$ is the actual BD position at time instance $k$, $\hat{\bm{p}}[k]=(\hat{p}_{x}[k],\,\hat{p}_{y}[k])$ is the estimated position, and $K$ is the total number of estimates. 
The MAE values in the $x$ and $y$ directions are given by
\begin{equation}
\bar{\epsilon}_{x} = \frac{1}{K} \sum_{k=1}^{K} \left| p_{x}[k]-\hat{p}_{x}[k] \right|, \quad\bar{\epsilon}_{y} = \frac{1}{K} \sum_{k=1}^{K} \left| p_{y}[k]-\hat{p}_{y}[k] \right|. 
\end{equation}
Four BD movement trajectories are simulated in the region: two straight lines labeled as paths 1 and 2 and two zigzag routes labeled as paths 3 and 4. 
Each trajectory is designed to cross the VLC cells and their overlapping areas to stress-test the tracking system. 
Since the height of the LED APs is fixed in the experimental setup, various vertical positions of the BD, ranging from 1.00 to 1.70~m, are simulated to examine the effect of the VLC cell size on the positioning performance. 
Each trajectory is simulated 500 times to collect the error statistics. 
As an example, Fig.~\ref{fig:all_sim_track} visualizes the simulated BD tracking results for the four trajectories with $h_{\textrm{BD}}=1.57$~m. 
The green lines indicate estimated trajectories over all simulation
repetitions, and the blue solid line shows the mean estimate.
The arrows indicate the directions of movement. The results show that the estimated trajectories closely follow the ground truth under the proposed algorithm.
Fig.~\ref{fig:all_sim} presents the overall positioning performance across all trajectories at different vertical positions of the BD.
More specifically, Fig.~\subref*{fig:sim_rmse_straight} and Fig.~\subref*{fig:sim_mae_straight} display the RMSE and MAE for the two straight paths, whereas Fig.~\subref*{fig:sim_rmse_zigzag} and Fig.~\subref*{fig:sim_mae_zigzag} display the results for the two zigzag paths.

The results indicate that, when the vertical distance between the LED APs and the BD decreases as $h_{\textrm{BD}}$ increases, the RMSE and MAE first decrease and then increase. 
The best performance is observed around $h_{\textrm{BD}}=1.57$~m, with RMSE values of about 0.227, 0.228, 0.220, and 0.228~m for the four paths, respectively.
The improvement at moderate heights comes from stronger LoS reception to the nearest LED AP, though if the BD is placed too high, overlapping regions between VLC cells diminish, creating outage zones where no VLC signal is received, as shown in Fig.~\ref{fig:sim_vl_power}.
These outage zones also illustrate the infrastructure requirement of the proposed indoor positioning system. Their size depends on the density of the luminaire grid, the overlap among VLC cells, and the installation geometry. Therefore, if the LED AP deployment becomes sparser, or if one LED AP fails without redundant overlap from nearby luminaires, the probability of dark zones increases and the continuity of proximity reporting degrades accordingly.
Moreover, the effect on the $y$-direction error is found to be more pronounced than that on the $x$-direction error. 
Fig.~\ref{fig:sim_cdf_all} shows the empirical cumulative distribution function (CDF) of tracking errors across all trajectories, indicating a median error of 0.220~m and a 90th-percentile error of 0.303~m, with errors ranging from 0.058~m to 0.878~m.
\subsection{Proof-of-Concept Experiments}\label{subsec:experiments}
\begin{figure*}
  \centering
  \hspace{0.1\columnwidth}
  \subfloat[]{\includegraphics[width=0.5\columnwidth]{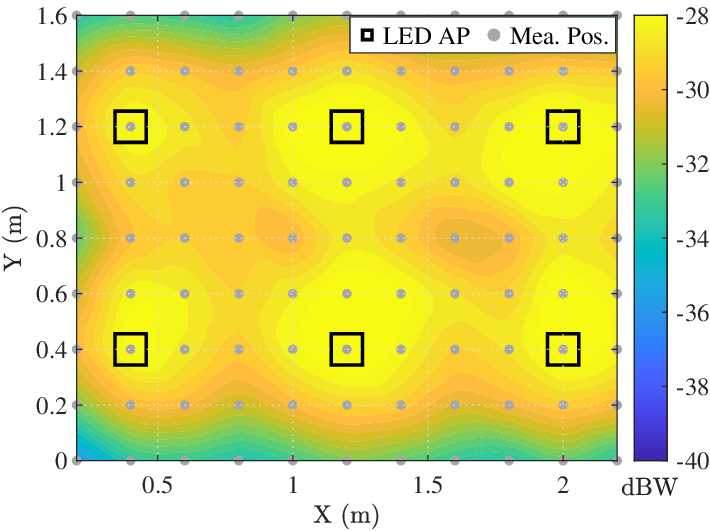}
   \label{fig:mea_vlc_power_h_140}}
  \hspace{0.1\columnwidth}
\subfloat[]{\includegraphics[width=0.5\columnwidth]{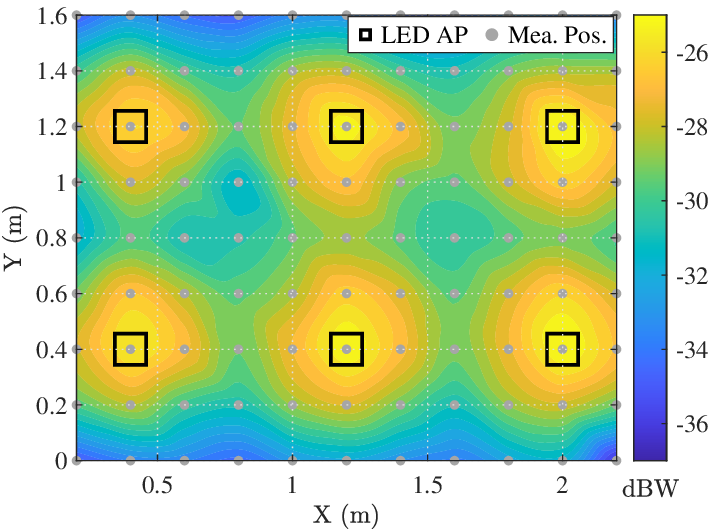}
   \label{fig:mea_vlc_power_h_157}}
  \hspace{0.1\columnwidth}
  \subfloat[]{\includegraphics[width=0.5\columnwidth]{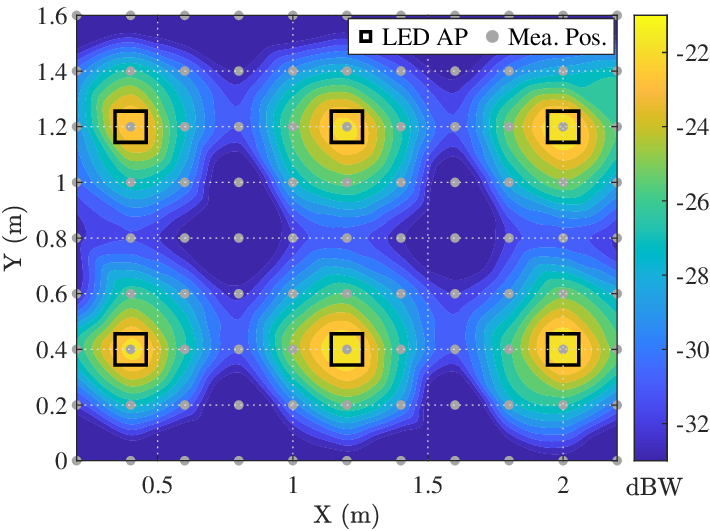}
   \label{fig:mea_vlc_power_h_170}}
  \hspace{0.1\columnwidth}
  \caption{Heatmap of the measured light intensity at the BD over the space of interest. 
  Colors range from dark blue to bright yellow, corresponding to the power levels of the received VLC signal.
  Different vertical distances between LED APs and the BD are demonstrated, showing different patterns of light reception:
  (a) $h_{\textrm{BD}} = 1.40$~m, (b) $h_{\textrm{BD}} = 1.57$~m, (c) $h_{\textrm{BD}} = 1.70$~m.
  }
  \label{fig:mea_vl_power}
\vspace{-8pt}
\end{figure*}
\begin{figure}
\vspace{-10pt}
  \centering
  \subfloat[]{\includegraphics[width=0.48\columnwidth]{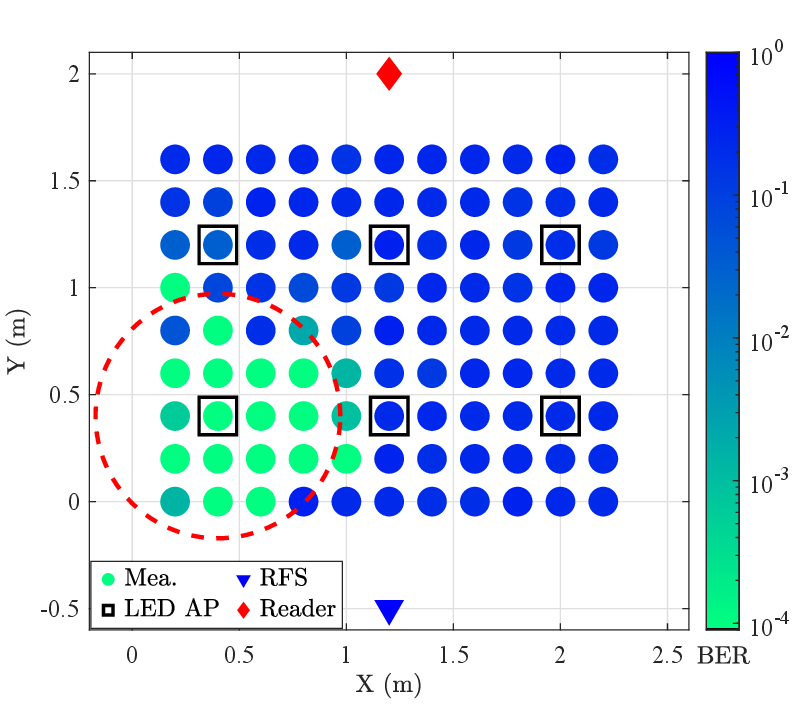}
   \label{fig:all_led_1_ber_spatial}}
  \hfill
  \subfloat[]{\includegraphics[width=0.48\columnwidth]{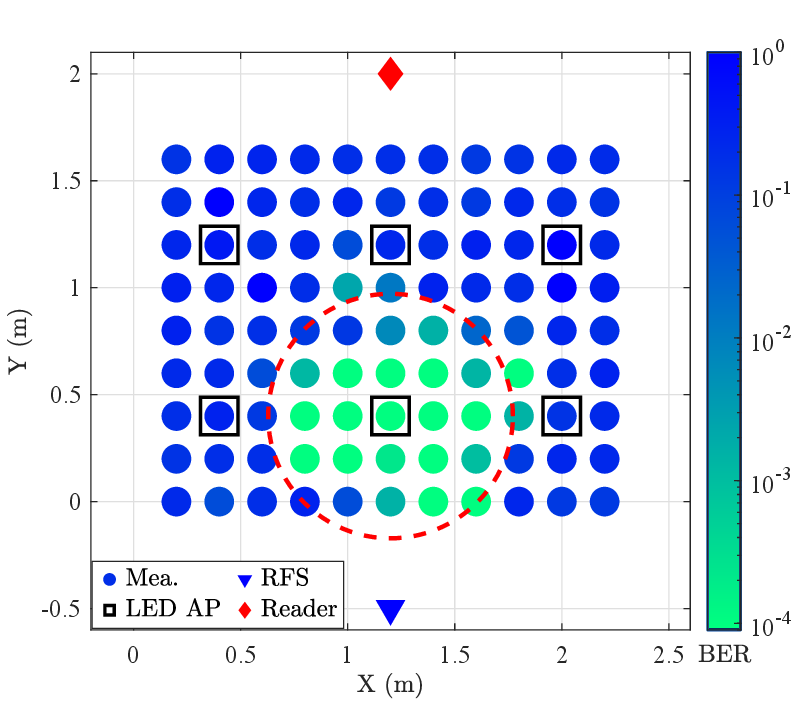}
   \label{fig:all_led_2_ber_spatial}}
  \hfill
  \subfloat[]{\includegraphics[width=0.48\columnwidth]{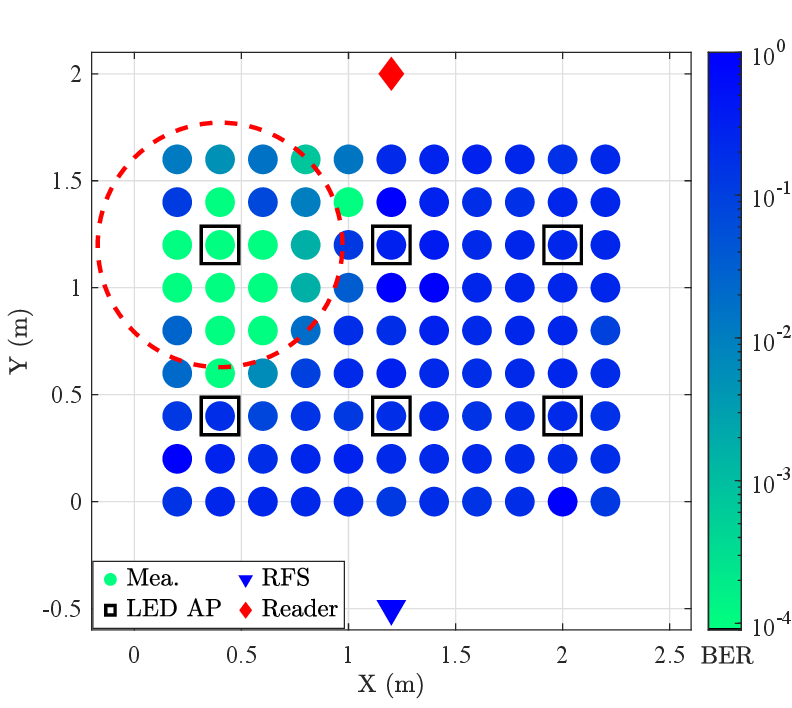}
   \label{fig:all_led_3_ber_spatial}}
  \hfill
  \subfloat[]{\includegraphics[width=0.48\columnwidth]{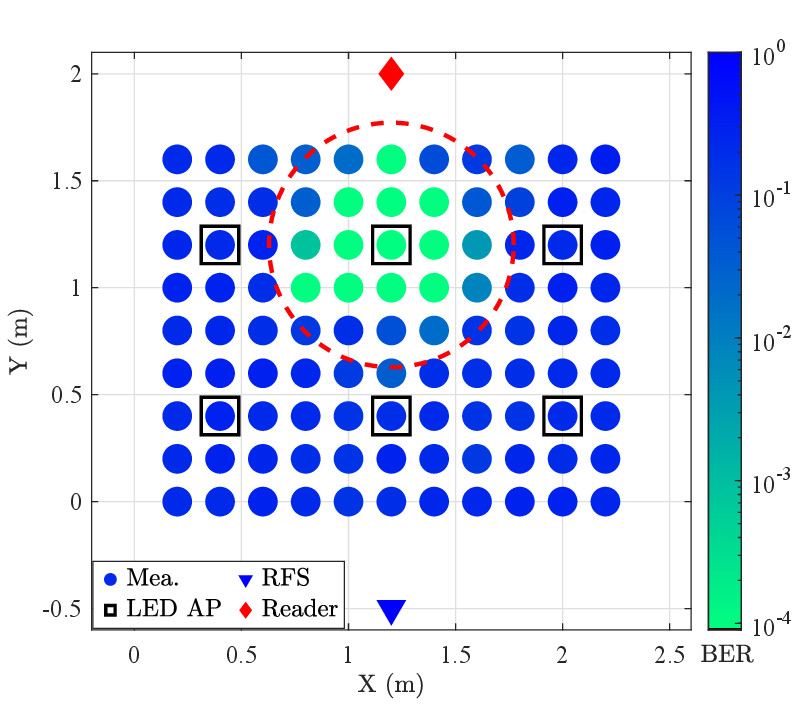}
   \label{fig:all_led_4_ber_spatial}}
  \hfill
  \subfloat[]{\includegraphics[width=0.48\columnwidth]{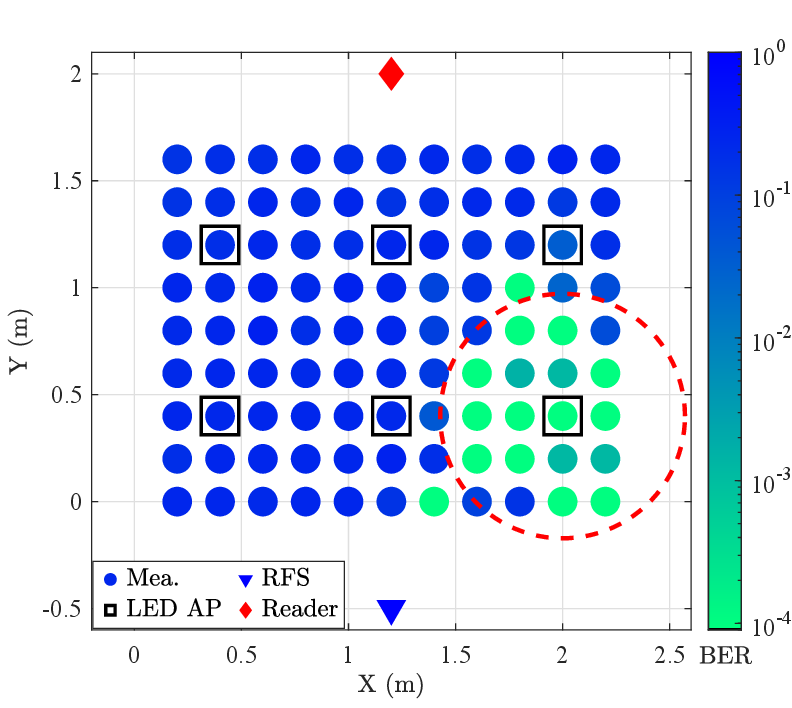}
   \label{fig:all_led_5_ber_spatial}}
  \hfill
  \subfloat[]{\includegraphics[width=0.48\columnwidth]{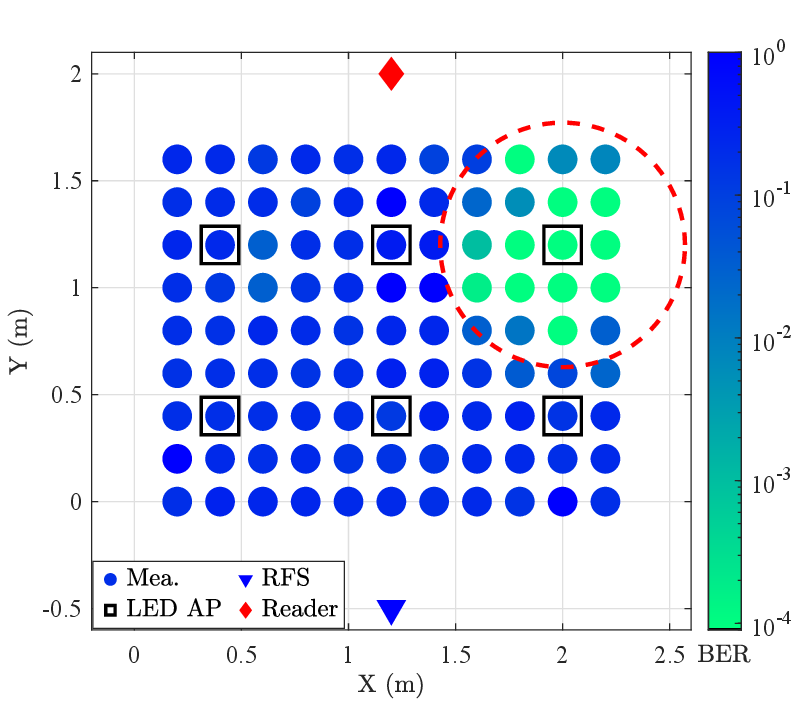}
   \label{fig:all_led_6_ber_spatial}}
  \hfill
  \caption{Spatial distribution of end-to-end communication BER measurements: (a)-(f) The decoded LED IDs originating from LED-1 through LED-6. 
  The red dashed circles indicate the estimated coverage range of the VLC cell with a radius $r_\textrm{cell} = 0.572$~m calculated by~\eqref{eq:vlc_cell_radius}. 
  Lower BER values denoted by lighter colored positions are observed inside the corresponding VLC cell, indicating the BD receives and forwards the LED ID belonging to that cell.
  }
  \label{fig:all_led_ber}
\vspace{-15pt}
\end{figure}
\begin{figure}
  \centering
  \subfloat[]
  {\includegraphics[width=0.48\columnwidth]{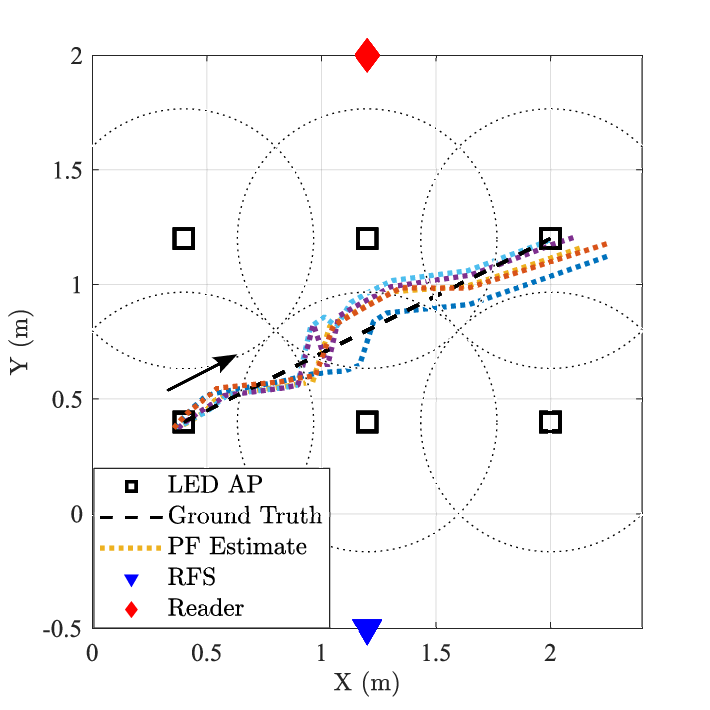}
   \label{fig:path_1_track_mea}}
  \hfill
  \subfloat[]
  {\includegraphics[width=0.48\columnwidth]{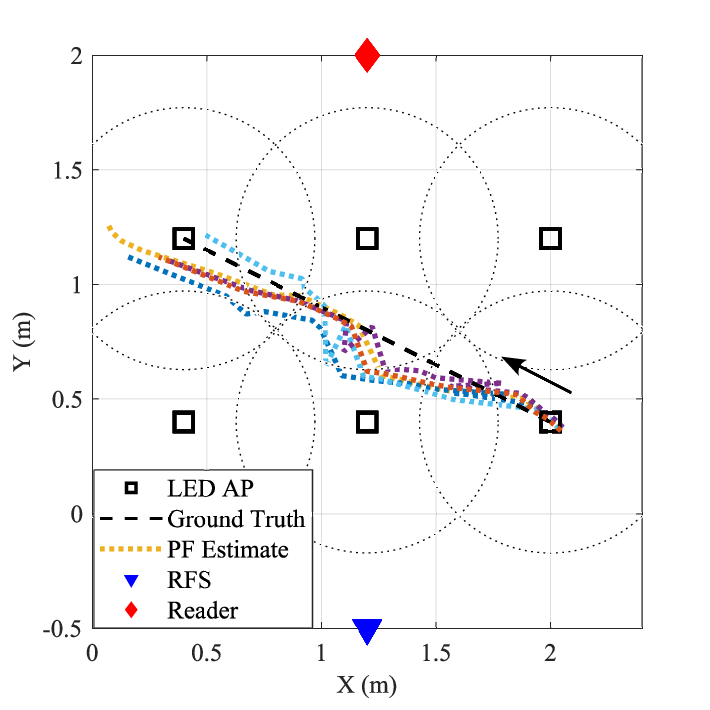}
   \label{fig:path_2_track_mea}}
  \hfill
  \subfloat[]
  {\includegraphics[width=0.48\columnwidth]{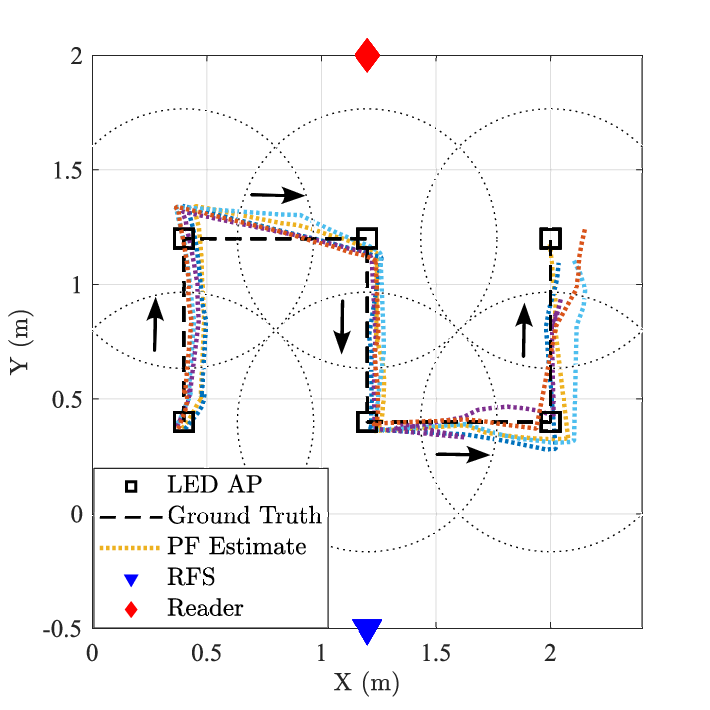}
   \label{fig:path_3_track_mea}}
  \hfill
  \subfloat[]
  {\includegraphics[width=0.48\columnwidth]{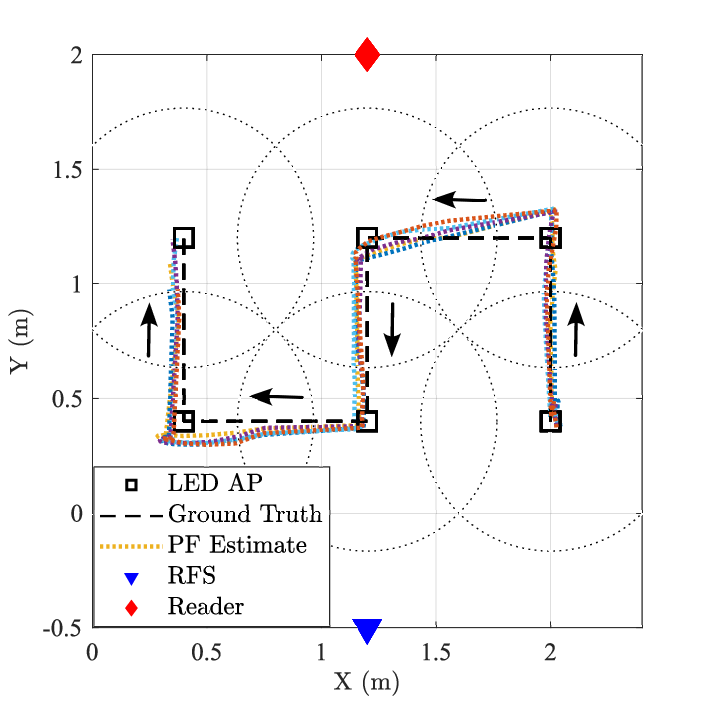}
   \label{fig:path_4_track_mea}}
  \hfill
  \caption{Experimental results of tracking the BD on paths 1-4 over five repetitions. The black dashed line is the ground truth of movement, the dotted lines are the estimated trajectories, and the arrow indicates the movement direction.}
  \label{fig:path_all_mea}
\vspace{-10pt}
\end{figure}
\begin{figure}
\centering
\includegraphics[width=0.9\columnwidth]{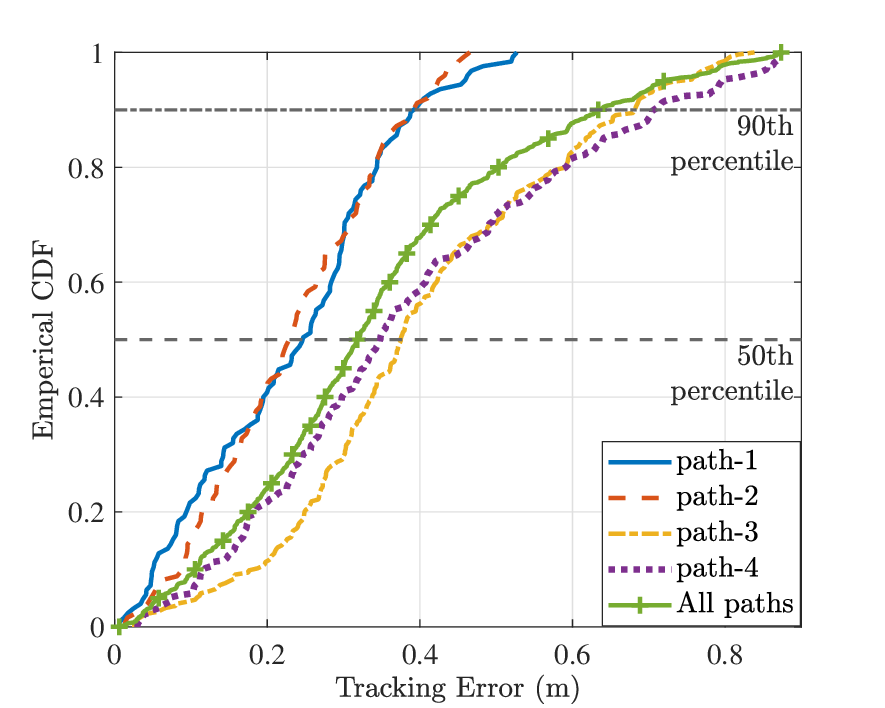}
\caption{Empirical CDF of the positioning errors from measurements across all trajectories.
For all tested paths, a median error of 0.318~m and a 90th-percentile error of 0.634~m are observed.}
\label{fig:cdf_mea_all}
\end{figure}
\begin{table*}
\caption{Experimental Results}
\resizebox{2\columnwidth}{!}{
\begin{tabular}{ccccccc|cccccc}
\cline{2-13}
       & \multicolumn{6}{c|}{Straight track}                                                                                                                                                            & \multicolumn{6}{c}{Zigzag track}                                                                                                                                                               \\
       & \multicolumn{3}{c|}{Path 1}                                                                              & \multicolumn{3}{c|}{Path 2}                                                         & \multicolumn{3}{c|}{Path 3}                                                                              & \multicolumn{3}{c}{Path 4}                                                          \\ \cline{2-13} 
       & RMSE~(m)  & $\overline{\epsilon_x} \pm \sigma_x$~(m) & \multicolumn{1}{c|}{$\overline{\epsilon_y} \pm \sigma_y$~(m)} & RMSE~(m)  & $\overline{\epsilon_x} \pm \sigma_x$~(m) & $\overline{\epsilon_y} \pm \sigma_y$~(m) & RMSE~(m)  & $\overline{\epsilon_x} \pm \sigma_x$~(m) & \multicolumn{1}{c|}{$\overline{\epsilon_y} \pm \sigma_y$~(m)} & RMSE~(m)  & $\overline{\epsilon_x} \pm \sigma_x$~(m) & $\overline{\epsilon_y} \pm \sigma_y$~(m) \\ \hline
Exp 1. & 0.282 & 0.218 ± 0.018                        & \multicolumn{1}{c|}{0.096 ± 0.005}                        & 0.236 & 0.151 ± 0.008                        & 0.144 ± 0.005                        & 0.431 & 0.186 ± 0.042                        & \multicolumn{1}{c|}{0.262 ± 0.042}                        & 0.468 & 0.167 ± 0.040                        & 0.305 ± 0.060                        \\
Exp 2. & 0.278 & 0.220 ± 0.012                        & \multicolumn{1}{c|}{0.115 ± 0.004}                        & 0.242 & 0.181 ± 0.009                        & 0.106 ± 0.006                        & 0.480 & 0.225 ± 0.057                        & \multicolumn{1}{c|}{0.280 ± 0.046}                        & 0.499 & 0.185 ± 0.047                        & 0.324 ± 0.065                        \\
Exp 3. & 0.287 & 0.229 ± 0.020                        & \multicolumn{1}{c|}{0.074 ± 0.006}                        & 0.273 & 0.196 ± 0.014                        & 0.134 ± 0.005                        & 0.432 & 0.192 ± 0.041                        & \multicolumn{1}{c|}{0.262 ± 0.042}                        & 0.397 & 0.134 ± 0.026                        & 0.253 ± 0.051                        \\
Exp 4. & 0.241 & 0.189 ± 0.014                        & \multicolumn{1}{c|}{0.065 ± 0.005}                        & 0.257 & 0.148 ± 0.016                        & 0.145 ± 0.008                        & 0.431 & 0.195 ± 0.045                        & \multicolumn{1}{c|}{0.255 ± 0.039}                        & 0.364 & 0.111 ± 0.019                        & 0.233 ± 0.048                        \\
Exp 5. & 0.244 & 0.206 ± 0.010                        & \multicolumn{1}{c|}{0.061 ± 0.004}                        & 0.264 & 0.190 ± 0.013                        & 0.128 ± 0.005                        & 0.438 & 0.226 ± 0.058                        & \multicolumn{1}{c|}{0.228 ± 0.033}                        & 0.481 & 0.148 ± 0.032                        & 0.333 ± 0.068                        \\ \hline
\end{tabular}
\label{tab:mea_result}}
\end{table*}
Proof-of-concept experiments are conducted in an indoor environment with parameters consistent with the ones that have been used in the simulations. 
Fig.~\ref{fig:implement_diagram} illustrates the hardware configuration with details discussed in Section~\ref{sec:implementations}. 
Light intensity measurements are conducted to characterize the illumination coverage in the region and to compare it with the simulation results presented in Fig.~\ref{fig:sim_vl_power}. 
A light meter is placed at 99 positions on a 0.2~m grid, which are marked with gray dots in Fig.~\ref{fig:mea_vl_power}.
The light intensity is measured at three different vertical positions of the BD for consistency with the simulations. 
Fig.~\ref{fig:mea_vl_power} visualizes the measured power heatmaps after interpolation. The results show that the spatial distribution of the received light power closely matches the simulated VLC cell coverage patterns in Fig.~\ref{fig:sim_vl_power}, including the approximately circular illumination zones and dark regions indicating potential outages where the BD cannot receive sufficient VLC signals to forward its proximity information.

To further evaluate the effective VLC reception region of the BD for each LED AP, end-to-end communication bit-error-rate (BER) measurements are conducted.
The BD is placed at the same 99 positions on a 0.2~m grid in the space at a height of $1.57$~m, and the reader USRP samples for 110 seconds at each marked position.
The recorded samples are processed using the workflow discussed in Section~\ref{subsec:signal_processing}, after which the BER is calculated.
Figs.~\subref*{fig:all_led_1_ber_spatial}-\subref*{fig:all_led_6_ber_spatial} show the spatial BER measurement results for the decoded LED IDs originating from LED-1 through LED-6, superimposed with red circles indicating the theoretical VLC cells with a radius $r_\textrm{cell} = 0.572$~m calculated using~\eqref{eq:vlc_cell_radius}. 
BER values below $10^{-3}$ are observed inside the corresponding VLC cells, indicating that the BD can receive and forward the LED IDs belonging to those cells. 
The overall results confirm end-to-end communication under the tested conditions between the LED APs, BD, RFS, and reader, allowing the BD to report its proximity within each relevant VLC cell.

To evaluate the tracking performance of the proposed system, four BD movement trajectories identical to those used in the simulations are tested.
The BD is affixed to a stand at a height of $1.57$~m, which is moved along the trajectories at approximately 0.36~m/s for the straight-line paths and 0.40~m/s for the zigzag paths. 
The experiment on each path is repeated five times.
During the experiment, the individual in charge of it carefully moves the BD while following a metronome to maintain a near-constant walking speed and maintaining a bent posture to avoid self-blocking of the optical path towards the LED APs. This procedure was intentionally adopted to create a controlled proof-of-concept evaluation under mostly unobstructed LoS conditions, so that the feasibility of the proposed VLC-BC localization principle could be evaluated without random human-body shadowing effects.
The position of the BD is estimated on the laptop connected to the USRP, which continuously samples and processes the received signals using the methodology discussed in Section~\ref{subsec:signal_processing} and the tracking algorithm discussed in Section~\ref{subsec:pf_tracking}.

Fig.~\ref{fig:path_all_mea} presents the tracking results of all tested trajectories, and Fig.~\ref{fig:cdf_mea_all} shows the empirical CDF of positioning errors.
The numerical results are summarized in Table~\ref{tab:mea_result}. For straight-line path 1, the average RMSE is 0.266~m, with a median error of 0.247~m and a 90th-percentile error of 0.396~m.
Similarly, for path 2, the average RMSE is 0.254~m, with a median error of 0.230~m and a 90th-percentile error of 0.393~m.
For zigzag path 3, the average RMSE is 0.442~m, with a median error of 0.375~m and a 90th-percentile error of 0.682~m.
Similarly, for path 4, the average RMSE is 0.442~m, with a median error of 0.349~m and a 90th-percentile error of 0.709~m.
The more complex movements on path-3 and path-4 produce larger positioning errors than the movements on path-1 and path-2.
This performance gap may partly result from the near-constant velocity motion model implemented by the PF, which is well-suited for straight-line movement, whereas the frequent heading changes characteristic of zigzag trajectories introduce greater uncertainty into the particle approximation.
Finally, across all the paths, a median positioning error of 0.318~m and a 90th-percentile error of 0.634~m are observed.
The results demonstrate submeter-level positioning accuracy with the low-complexity, batteryless BD in the tested setup.

A comparison of Fig.~\ref{fig:sim_cdf_all} and Fig.~\ref{fig:cdf_mea_all} shows that the measured CDF has larger errors than the simulated CDF, and the discrepancy is particularly pronounced at the 90th percentile. 
This behavior suggests that the nominal operating point of the simulation is broadly consistent with the experiment, but the real prototype experiences occasional adverse events that are not fully represented in the simulation setup based on simplified models.
Several practical factors contribute to these larger tail errors. 
First, the LED illumination patterns and VLC coverage are imperfect in practice, as shown by the comparison of Fig.~\ref{fig:sim_vl_power} and Fig.~\ref{fig:mea_vl_power}. 
Although the BER measurements confirm low BER inside the corresponding VLC cells, occasional missed or delayed LED-ID detections may still occur near cell-edge regions, transitions between overlapping VLC cells, or during temporary weak illumination. 
Second, the PF uses a simplified kinematic model, while the experimental motion is performed manually and therefore includes small irregularities in speed, orientation, and heading. 
This effect is visible in the zigzag trajectories, where the measured errors are larger than in straight-line cases. 
Third, the real-world backscatter link is affected by indoor multipath, hardware imperfections, and signal-processing imperfections that are insufficiently captured by the empirical channel model used in the simulations.
%
\begin{figure}[!t]
\centering
\includegraphics[width=0.98\columnwidth]{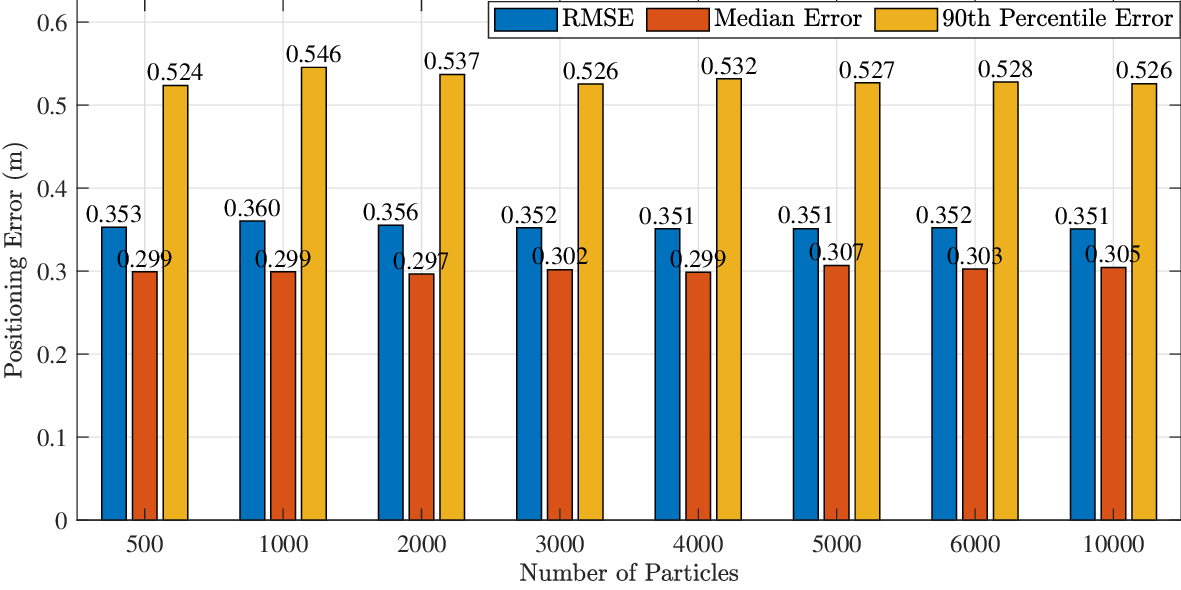}
\caption{Sensitivity analysis of tracking performance averaged over all four trajectories: positioning error versus the number of particles $N_\textrm{p}$.}
\label{fig:sensitivity_particles}
\end{figure}
\begin{figure}[!t]
\centering
\includegraphics[width=0.98\columnwidth]{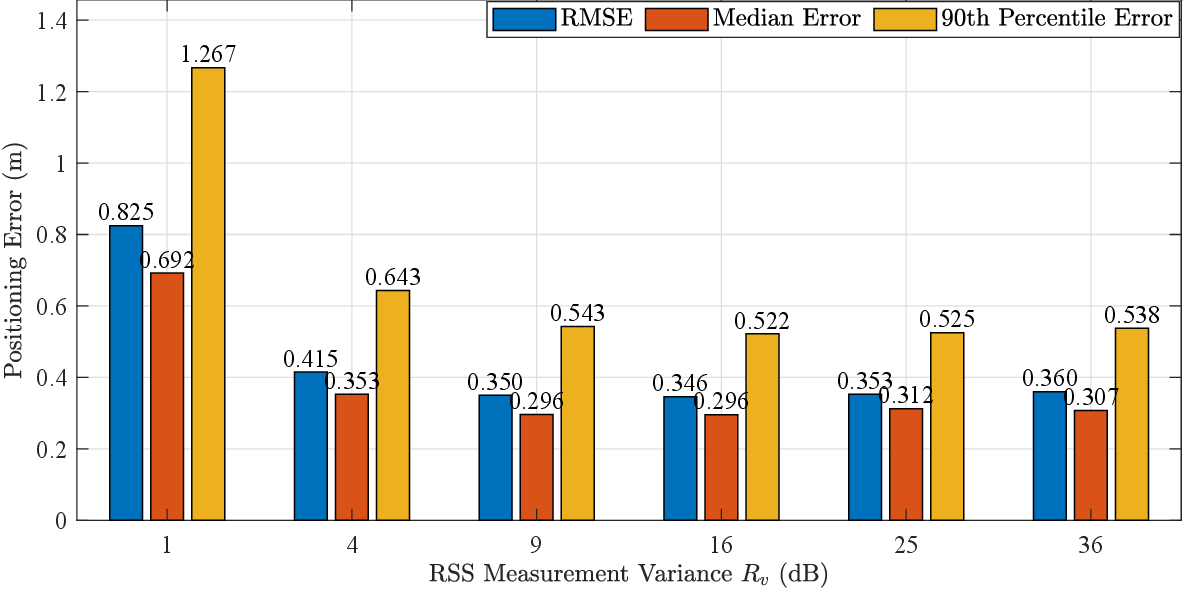}
\caption{Sensitivity analysis of tracking performance averaged over all four trajectories: positioning error versus the RSS measurement-noise variance $R_v$.}
\label{fig:sensitivity_rv}
\end{figure}

In addition to the default parameter configuration listed in Table~\ref{tab:parameters}, we further evaluate the sensitivity of the tracking algorithm to two key parameters, namely the number of particles $N_\textrm{p}$ and the RSS measurement-noise variance $R_v$.
Two figures show the resulting error metrics averaged over all trajectories. 
In Fig.~\ref{fig:sensitivity_particles}, the number of particles is varied as $N_\textrm{p}\in\{500,1000,2000,3000,4000,5000,6000,10000\}$. The RMSE remains in the range of 0.351-0.360~m, the median error remains in the range of 0.290-0.307~m, and the 90th-percentile error remains in the range of 0.524-0.546~m. Therefore, it is possible to conclude that increasing $N_\textrm{p}$ does not yield a consistent improvement in the considered setting. This indicates that, with the available LED-ID and RSS measurements, even the lower tested particle counts provide a sufficient approximation of the posterior distribution. Since the complexity scales approximately linearly with $N_\textrm{p}$, this result indicates that the particle number can be reduced when computational load is critical. 

Fig.~\ref{fig:sensitivity_rv} presents the sensitivity to the RSS measurement-noise variance, where $R_v$ is varied over $\{1,4,9,16,25,36\}$. The corresponding RSS standard deviation values are $\sigma_v\in\{1,2,3,4,5,6\}$~dB in the RSS domain. 
The performance degrades when $R_v=1$, where the RMSE, median error, and 90th-percentile error are 0.825~m, 0.692~m, and 1.267~m, respectively. 
This result indicates that an underestimated RSS variance makes the PF overconfident in the RSS model, causing the filter to overreact to RSS fluctuations or model mismatch. 
When $R_v$ is increased to 4, the three metrics decrease to 0.415~m, 0.353~m, and 0.643~m, respectively. For $R_v\in\{9,16,25,36\}$, the performance becomes stable, with RMSE between 0.346~m and 0.360~m, median error between 0.296~m and 0.312~m, and 90th-percentile error between 0.522~m and 0.543~m. 

These sensitivity studies show that the proposed PF tracking architecture is relatively insensitive to particle count within the tested range once a sufficient number of particles is used. They also show that the RSS variance should not be set too small, since overconfident RSS weighting can degrade tracking accuracy. A moderate-to-large $R_v$ provides a better balance between the continuous RSS observation, the coarse LED-ID proximity constraint, and the motion model.

\begin{table*}[!ht]
\centering
\small
\caption{Comparison with state-of-the-art VLP systems}
\label{tab:sota}
\resizebox{2\columnwidth}{!}{
\begin{tabular}{lllllll}
\hline
\textbf{System} & \textbf{Technique} & \textbf{Positioning Method} & \textbf{Representative Device-Side Supply/Platform} & \textbf{Hardware Complexity} & \textbf{Device Cost} & \textbf{Accuracy} \\ \hline
\cite{zhu2023perspective} & VLC (camera) & Image features & Battery & High & High & 0.05~m (50\%), 0.1~m (90\%) \\
\cite{epsilon} & VLC (camera) + IMU & Proximity + RSS trilateration & Battery & High & High & 0.4~m (50\%), 0.8~m (90\%) \\
\cite{xie2018vlp_proximity} & VLC (camera) & Proximity & DC power supply & High & High & Approx. 0.35~m \\
\cite{yang2023vlp_advance} & VLC (APD) & Phase & Powered embedded platform (Raspberry Pi-based) & High & High & 0.043~m (50\%), 0.098~m (90\%) \\
\cite{cherntanomwong2015prox} & VLC (PD) & Proximity & Powered embedded platform (Arduino-based) & Medium & Medium & Approx. 1.5~m \\
\cite{albraheem2023vlc_ble} & VLC (PD) + Bluetooth & Proximity + RSS trilateration & Battery & High & High & 0.03-0.52~m \\
\cite{cappelli2022selfsufficient} &
VLC (PV module) + LoRaWAN &
RSS lateration &
Energy harvesting + rechargeable battery/MCU &
Medium &
Medium &
0.0128~m (50\%) \\

\cite{perera2025mlaided} &
VLC/LIoT (PV EH + OSD) + IR uplink &
ML/DNN-based positioning &
Battery-free intermittent PV EH / embedded prototype &
Medium &
Medium &
80\% within 0.125~m tolerance \\

\cite{shao2018retro} &
VLC (retroreflector + LCD shutter) &
RSS/trilateration &
Ultra-low-power optical tag &
Medium &
Medium &
0.02~m (90\%) \\

\cite{shao2019passiveretro} &
VLC (passive retroreflector) &
RSS/trilateration &
Completely passive optical tag &
Low &
Low &
0.065~m (90\%) \\

\cite{nazzal2022retrovlp} &
VLC (retroreflector + PD array) &
Reflected-power centroid + height variance &
Passive retroreflector/no active tag electronics &
Low &
Low &
cm- to dm-level \\

\textbf{This work} & VLC (PV cell) + BC & Proximity + RSS model & \textbf{Device-side energy harvesting} & \textbf{Low} & \textbf{Low} & 0.318~m (50\%), 0.634~m (90\%) \\ \hline
\end{tabular}}
\vspace{-10pt}
\end{table*}
\subsection{Comparison with State-of-the-Art Systems}
For comparison purposes, Table~\ref{tab:sota} contrasts our joint VLC-BC design against representative indoor VLP and hybrid localization baselines in terms of technology, positioning method, representative device-side energy supply/platform, hardware complexity, device cost, and achieved accuracy. 
Although the comparison is carried out among papers that consider different assumptions that are not strictly controlled, it helps position the proposed method within the existing literature.
Nevertheless, Table~\ref{tab:sota} does not provide a complete infrastructure-power comparison, since most referenced works do not report end-to-end infrastructure power in a directly comparable manner. 
Therefore, the table mainly positions the proposed system in terms of tag-side energy supply, device complexity, and representative localization performance, rather than as a full system-level energy-budget dimensioning.
Three trends emerge in the surveyed systems. 
First, camera- and advanced photodetector-based designs can achieve centimeter-level to decimeter-level accuracy, whereas they depend on power-intensive photonic front ends and substantial on-board signal processing, resulting in higher device-side complexity and cost~\cite{zhu2023perspective,epsilon,yang2023vlp_advance,xie2018vlp_proximity}. 
Second, lower-complexity proximity-based or self-powered VLC systems reduce the burden at the tracked device, but they typically provide coarser location information or rely on simpler measurements.
Third, high-complexity, self-powered VLP systems in the literature adopt on-node position inference with active radio or optical transmission of the location information, making the nodes more complex and expensive~\cite{cappelli2022selfsufficient,perera2025mlaided,zhou2025resource}. Nevertheless, they can provide better positioning accuracy. 

Our design differs from conventional VLP systems and prior VLC-BC communication prototypes in its objective and division of functions. Conventional VLP systems mainly pursue higher positioning accuracy with complex optical receivers or active companion radios, whereas prior VLC-BC studies primarily demonstrate communication bridging between light and RF domains. On the other hand, the present work targets localization with an energy-neutral node as the main design objective. VLC is used only for low-overhead cell identification, BC is used only for passive proximity reporting, and the fusion and tracking tasks are shifted to the reader side. In the tested setup, this division of functions provides submeter-level tracking without cameras, TIAs, active RF transmitters, or on-device localization processing, thereby distinguishing the proposed system from both existing localization-oriented VLP solutions and communication-oriented VLC-BC systems.

In the experiments, the proposed VLC-BC design achieves a median positioning error of 0.318~m and a 90th-percentile error of 0.634~m, which is below the lower end of the indoor tracking accuracy range specified for 3GPP Ambient IoT, where TS~22.369 lists a required positioning accuracy of 1-3~m with 90\% availability for indoor tracking scenarios~\cite{3gpp22369}.
The positioning errors are larger than those reported by centimeter-level camera- or APD-based systems, yet this gap is an expected consequence of the measurement and hardware choices made in the present architecture. The proposed BD does not capture rich optical observables such as image features, precise optical RSS, or optical phase. Instead, the reader estimates the position from two comparatively coarse observables: a set of detected LED-cell IDs that provides proximity-level spatial constraints and the RSS of the backscatter signal that refines the estimate but is affected by RF multipath propagation and fading.
Referring to the application granularity reported in recent indoor positioning surveys, the proposed system belongs to the submeter or furniture-level class, rather than the component-level class that requires accuracy below 10~cm~\cite{sartayeva2023survey}.
Therefore, it may be suitable for coarse-to-intermediate asset awareness, such as trolley tracking, tool monitoring, equipment localization, and zone-level warehouse visibility. 
However, precision robot docking, automatic pickup, and manipulation-assisted alignment require higher accuracy, whose implementation costs can be better justified by the costs of the machinery that needs to be tracked~\cite{li2024indoor,huang2023indoor}.

\noindent\textit{Key Takeaways:}
While camera- and advanced PD-based VLP systems can provide higher-accuracy localization, their power and hardware demands make them less suitable for large-scale, battery-free asset-tracking deployments. 
On the other hand, the proposed VLC-BC system preserves the energy-neutral, low-profile form factor while providing submeter-level accuracy in the tested setup, making it a potential option for asset-tracking applications in which development and maintenance costs are primary considerations.

\subsection{Discussion}
The results indicate that the proposed VLC-BC architecture can support submeter-level asset tracking in the setup under evaluation in this paper while keeping the BD energy-neutral and hardware-efficient. 
Instead of performing sensing, localization, and RF transmission at the IoT node, the BD only harvests light energy, receives LED IDs, and backscatters the corresponding proximity information. The more complex signal processing and PF-based data fusion are shifted to the edge reader, which reduces device-side processing and power requirements for battery-free indoor tracking.
From an application perspective, the results suggest that the current submeter accuracy may support furniture-level or zone-level asset tracking, including equipment monitoring, cart tracking, and coarse mobile-asset visibility~\cite{sartayeva2023survey}. Component-level robotic tasks, such as precise docking and manipulation-assisted alignment, require tighter accuracy and remain outside the scope of the current design.

However, several practical implementation issues should be considered when deploying the system in actual environments. First, the VLC proximity information depends strongly on the illumination geometry. The size and overlap of VLC cells are affected by LED spacing, mounting height, BD height, receiver field of view, and BD orientation. If the overlap between adjacent cells is insufficient, outage service regions may appear where no LED ID can be decoded. Conversely, excessive overlap may increase ambiguity in proximity reports. A practical deployment should therefore jointly optimize LED placement, frequency reuse, and VLC-cell overlap. In addition, orientation-aware cell modeling or calibration-based coverage maps could be incorporated at the reader to compensate for tilted BDs and non-ideal illumination patterns without increasing the BD complexity.

Second, the backscatter RSS measurement is sensitive to multipath fading, human blockage, antenna polarization, and the relative placement of the RFS, BD, and reader. These factors may introduce deviations between the theoretical RSS model and the measured signal strength, especially in cluttered indoor spaces. A possible solution is to employ multiple RF readers or multiple RFSs to provide spatial diversity and reduce the dependence on a single BC link. Alternatively, site-specific RSS calibration, adaptive RSS weighting, or fingerprint-assisted correction could be integrated into the measurement likelihood of the PF to reduce sensitivity to environmental variations.

Third, the current PF adopts a near-constant velocity motion model, which performs well for smooth trajectories yet becomes less accurate when the BD experiences abrupt turns or irregular motion, as observed in the zigzag-path experiments. This limitation can be mitigated by using adaptive motion models, interacting multiple-model filtering, or learning-assisted motion prediction at reader side. These approaches would preserve the lightweight BD design because the additional computation would be performed only at the edge.

Finally, scalability should be considered for larger deployments. Although the FDM-based VLC design avoids the need of a strict synchronization in the time domain, the number of available frequency pairs is finite and must be planned carefully for reuse in dense LED networks. Graph-based frequency planning, hierarchical LED-ID assignment, and frequency-reuse patterns in VLC clusters can be adopted to effectively mitigate co-channel interference. Moreover, the energy-neutral operation of the BD requires sufficient illumination; in dim or intermittently blocked line-of-sight propagation conditions, adding a small storage capacitor and duty-cycled backscatter operation in the BD can improve reliability. These considerations indicate that the proposed system is most suitable for indoor asset tracking scenarios in which lighting infrastructure is available, submeter-level accuracy is sufficient, and low maintenance cost is a primary requirement.
%
\section{Conclusion}\label{sec:conclusion}
This paper presented a joint VLC-BC system for proximity-based indoor asset tracking with energy-neutral IoT devices. 
The proposed BD harvests energy from LED illumination, receives LED IDs through VLC, and forwards the proximity information by modulating and backscattering ambient RF carriers, thereby avoiding an active RF synthesizer and complex photonic front ends. A multi-cell FDM-based VLC deployment was designed to distinguish adjacent LED APs, and a lightweight PF algorithm was developed at the edge reader to fuse decoded LED IDs with BC RSS measurements. Simulations and proof-of-concept experiments demonstrated the operation of the complete light-to-RF tracking chain in the tested setup. The experimental results yielded a median positioning error of 0.318~m and a 90th-percentile error of 0.634~m across multiple trajectories. These results demonstrate submeter-level tracking in the tested setup while retaining a simple, energy-neutral BD; validation in larger and more varied deployments remains future work.

\bibliographystyle{IEEEtran}
\bibliography{reference}

\end{document}